\documentclass[pre,aps,onecolumn,superscriptaddress,nofootinbib]{revtex4-1}
\usepackage{amssymb}
\usepackage{amsmath}
\usepackage{graphicx}
\usepackage{array}
\usepackage{mathtools}
\usepackage{amssymb}
\usepackage{verbatim}
\usepackage{color}
\usepackage{bm}
\usepackage{amsthm}
\usepackage[normalem]{ulem}
\usepackage[titletoc,title]{appendix}
\usepackage[colorlinks=true, urlcolor=blue, anchorcolor=blue, citecolor=blue,filecolor=blue,linkcolor=blue,menucolor=blue%,pagecolor=blue
]{hyperref}

\usepackage{color}
\definecolor{Blue}{rgb}{0.00, 0.00, 0.80}
\definecolor{Red}{rgb}{0.80, 0.00, 0.00}
\definecolor{Magenta}{rgb}{0.70, 0.00, 0.70}
\definecolor{Green}{rgb}{0.00, 0.50, 0.00}
\newcommand{\bea}{\begin{eqnarray}}
\newcommand{\eea}{\end{eqnarray}}
\newcommand{\be}{\begin{equation}}
\newcommand{\ee}{\end{equation}}
\newcommand{\nn}{\nonumber}
\newcommand{\bee}{\begin{equation*}}
\newcommand{\eee}{\end{equation*}}

\def\XXint#1#2#3{{\setbox0=\hbox{$#1{#2#3}{\int}$}
     \vcenter{\hbox{$#2#3$}}\kern-.5\wd0}}

\def\XXint#1#2#3{{\setbox0=\hbox{$#1{#2#3}{\int}$}
     \vcenter{\hbox{$#2#3$}}\kern-.5\wd0}}

\usepackage{dsfont}
\usepackage{fixmath}
\begin{document}

\title{Subleading-order theory for condensation transitions in large deviations of sums of independent and identically distributed random variables}

\author{Naftali R. Smith}
\email{naftalismith@gmail.com}
\affiliation{Racah Institute of Physics, Hebrew University of Jerusalem, Jerusalem 91904, Israel}

%\pacs{05.40.-a, 05.70.Np, 68.35.Ct}

\begin{abstract}

We study the full distribution $P_{N}\left(A\right)$ of sums 
$A = \sum_{i=1}^N$
where $x_1, \dots, x_N$ are $N \gg 1$ independent and identically distributed  random variables each sampled from a given distribution $p(x)$ with a subexponential $x \to \infty$ tail. We consider two particular cases:
(I) the one-sided stretched exponential distribution $p(x) \propto e^{-x^\alpha}$ where $0 < x < \infty$,
(II) the two-sided stretched exponential distribution $p(x) \propto e^{-|x|^\alpha}$ where $-\infty < x < \infty$.
We assume $0 < \alpha < 1$ (in both cases).
As follows immediately from known theorems, for both cases (i) typical fluctuations of $\Delta A = A - \left\langle A\right\rangle $ are described by the central-limit theorem, (ii) the tail $A \to \infty$ is described by the big-jump principle $P_{N}\left(A\right) \simeq N p\left(A\right)$, and (iii) in between these two regimes there is a nontrivial intermediate regime which displays anomalous scaling
$P_{N}\left(A\right) \sim e^{-N^\beta f(\Delta A/N^\gamma)}$
with anomalous exponents $\beta,\gamma \in (0,1)$ and large-deviation function $f(y)$ that are all exactly known. 
In practice, although these theoretical predictions of $P_{N}\left(A\right)$ work very well in regimes (i) and (ii), they often perform quite poorly in the intermediate regime (ii), with errors of several orders of magnitude for $N$ as large as $10^4$.
We calculate subleading order corrections to the theoretical predictions in the intermediate regime. We find that for $0 < \alpha < \alpha_c$, these corrections scale as power laws in $N$, while for $\alpha_c < \alpha < 1$ they scale as stretched exponentials, where the threshold value is $\alpha_c = 1/2$ in case (I) and $\alpha_c = 2/3$ in case (II). This difference between the two cases is a result of the mirror symmetry $p(x) = p(-x)$ which holds only in the latter case.

\end{abstract}

\maketitle

{%
        %\singlespacing
        \hypersetup{linkcolor=blue}
        \tableofcontents
}

\section{Introduction}

The study of distributions $P_{N}\left(A\right)$ of sums $A = x_1 + \dots + x_N$ of $N$ independent and identically distributed (i.i.d.) random variables $x_1  ,  \dots  , x_N$, each sampled from a given distribution $p(x)$, is of central, fundamental importance for
 a large number of scientific fields, including statistical physics and probability theory \cite{Petrov1975}, with applications ranging from random walks \cite{Denisov08, Barkai20} to active matter \cite{GM19, MLMS21, MGM21} and more.
Much is known regarding the behavior of such distributions in the limit $N \gg 1$.
Typical fluctuations of $A$ are described by the central limit theorem (CLT) provided the first and\\
 second cumulants of the distribution $p(x)$, i.e., the mean $\mu=\left\langle x_{1}\right\rangle $ and the variance $\sigma^2 = \text{Var}(x_1)$, are both finite. %(otherwise, they are described by L\'{e}vy distributions).
However, large deviations of $A$ are important to study as well despite their rarity, because they can represent dramatic
 and sometimes catastrophic events %with significant consequences
in realistic systems,
e.g. droughts, floods, heatwaves or cold spells \cite{AS24}.
One of the most important problems about such events is to estimate their likelihood. 
More generally, large deviations (or rare\\ events) attract interest at a more fundamental level, e.g., because their study can improve our understanding of nonequilibrium statistical mechanics \cite{Hugo2009, MS2017, Touchette2018}.
Large deviations of sums of i.i.d. random variables are generally described either by the large deviations principle (LDP) as follows from Cram\'{e}r's theorem \cite{Cramer38} if the tail of $p(x)$ decays superexponentially, or by the big-jump principle (BJP) \cite{Chistyakov64,Foss13,Denisov08,Geluk09,Clusel06, BCV10,BUV14,VBB19, WVBB19,Gradenigo13,Barkai20, MKB98, BBBJ2000, EH05, MEZ05, EMZ06, Majumdar10, CC12, ZCG, Corberi15, BVB24, HBB24, VB24, BVBB25} if the  tail of $p(x)$ decays subexponentially (the marginal case, in which the decay is exponential, is also of interest and has been studied, see e.g. Ref.~\cite{MLMS21}).

If the tail $x \to \infty$ of $p(x)$ decays as a stretched exponential $- \ln p(x) \sim x^\alpha$ with $0 < \alpha < 1$, an interesting intermediate regime of $P_{N}\left(A\right)$ emerges between the Gaussian distribution of typical fluctuations of $A$ and the stretched exponential behavior of $P_{N}\left(A\right)$ at $A \to \infty$ \cite{EMZ06, Majumdar10, GM19, BKLP20, GIL21, GIL21b,MLMS21, MGM21, IP25}. In the intermediate regime, $P_{N}\left(A\right)$ is described by an anomalous LDP whose associated rate function has a corner singularity at which its first derivative jumps. This singularity is often interpreted as a ``condensation transition'', which separates between a subcritical homogeneous phase in which $x_1, \dots, x_N$ are all of the same order of magnitude, and a supercritical condensed phase in which one of the $x_i$'s is of order $A$, while the rest of the $x_i$'s are of the same order magnitude as each other.

In practice, one often finds that, while the CLT, LDP and BJP are valid at moderately large values of $N$ ($N \simeq 30$ is usually sufficient), the leading-order theory in the intermediate regime only gives a good description of $P_{N}\left(A\right)$ at extremely large $N$'s, of order $N \sim 10^5$ or even larger \cite{GM19, MGM21}. As a result, simple, reliable predictions for $P_{N}\left(A\right)$ in the intermediate regime are not available, and one must resort to numerical computations. 
In this paper we aim to fill this gap by developing a subleading-order $N \gg 1$ theory for the intermediate regime. In our calculations, we use the two particular examples $p(x) = p_i(x)$ of the one-sided and two-sided stretched-exponential distributions,
\bea
\label{SE1}
p_{1}\left(x\right)&=&\frac{e^{-x^{\alpha}}}{\Gamma\left(1+\frac{1}{\alpha}\right)},\quad0<x<\infty,\\[1mm]
\label{SE2}
p_{2}\left(x\right)&=&\frac{e^{-\left|x\right|^{\alpha}}}{2\Gamma\left(1+\frac{1}{\alpha}\right)},\quad-\infty<x<\infty
\eea
(where $\Gamma$ is the gamma function) as prototypical examples for asymmetric ($p(x) \ne p(-x)$) and symmetric ($p(x) = p(-x)$) distributions respectively, and we briefly outline how to extend the results to general $p(x)$ with stretched-exponential tails. Although our main interest is the case $0 < \alpha < 1$ (corresponding to a subexponential $x \to \infty$ tail), for pedagogical purposes we also briefly address the case $\alpha > 1$ in which the $x \to \infty$ tail is superexponential.

The remainder of the paper is organized as follows. In Section \ref{sec:leading}, we give an overview of the standard leading-order $N\gg1$ results for $P_{N}\left(A\right)$ in the different regimes, without giving detailed derivations. In Sections \ref{sec:leadingDerivation} and \ref{sec:subleading} we derive the leading- and subleading-order $N\gg1$ results, respectively, for the subexponential case $0<\alpha<1$ in the intermediate regime, and test their performance. We also briefly discuss how to extend the results to general $p(x)$ with subexponential tails. In Section \ref{sec:Discussion} we briefly summarize and discuss our main findings.

\section{Leading-order $N \gg 1$ behavior: General overview}
\label{sec:leading}

\subsection{Typical fluctuations: Central Limit Theorem}
Using that
\be
\int_{0}^{\infty}x^{b}e^{-x^{\alpha}}dx=\frac{1}{\alpha}\Gamma\left(\frac{1+b}{\alpha}\right)
\ee
One finds that the means and variances of the stretched-exponential distributions \eqref{SE1} and \eqref{SE2} and  are given by
\bea
\mu_{1}&=&\int_{0}^{\infty}xp_{1}\left(x\right)dx=\frac{\Gamma\left(1+\frac{2}{\alpha}\right)}{2\Gamma\left(1+\frac{1}{\alpha}\right)}\,,\\
\mu_{2}&=&\int_{-\infty}^{\infty}xp_{2}\left(x\right)dx=0\,,\\
\sigma_{1}^{2}&=&\int_{0}^{\infty}x^{2}p_{1}\left(x\right)dx-\mu_{1}^{2} =\frac{4\Gamma\left(1+\frac{3}{\alpha}\right)\Gamma\left(1+\frac{1}{\alpha}\right)-3\Gamma\left(1+\frac{2}{\alpha}\right)^{2}}{12\Gamma\left(1+\frac{1}{\alpha}\right)^{2}} \,,\\
\sigma_{2}^{2}&=&\int_{-\infty}^{\infty}x^{2}p_{2}\left(x\right)dx=\frac{\Gamma\left(1+\frac{3}{\alpha}\right)}{3\Gamma\left(1+\frac{1}{\alpha}\right)} \, ,
\eea
respectively ($\mu_2$ vanishes due to the mirror symmetry $p_2(x) = p_2(-x)$).
According to the central limit theorem, typical fluctuations of $A$ are given by a Gaussian distribution
\be
\label{Gaussian}
P_{N}\left(A\right)\simeq P_{N}^{\left(\text{Gauss}\right)}\left(A\right) \equiv \frac{1}{\sqrt{2\pi N\sigma^{2}}}e^{-\left(A-N\mu\right)^{2}/\left(2N\sigma^{2}\right)}
\ee
with mean $\left\langle A\right\rangle  = N\mu$ and variance $\text{Var}\left(A\right) = N\sigma^2$. 
As a rule of thumb, for most distributions $p(x)$, the Gaussian approximation describes typical fluctuations accurately if $N \ge 30$. Indeed, in Fig.~\ref{figAlphaThreeHalves} we observe excellent agreement between this approximation \eqref{Gaussian} and the exact $P_{N}\left(A\right)$, which was computed by a numerical evaluation of the $N$th convolution power of $p_i(x)$, for $i=1,2$, $N=31$ and $\alpha=3/2$.
Incidentally, if better accuracy is required, one can obtain small corrections to the Gaussian approximation \eqref{Gaussian} by employing the Edgeworth expansion \cite{Edgeworth1905, Edgeworth1906, Kendall1948, EdgeworthWiki} (not shown).

In general, large deviations of $A$ significantly depart from the central limit theorem, and their behavior depends on the type of decay of the tails of $p(x)$ as we now describe.

\subsection{Large deviations for superexponential decay $\alpha>1$: Cram\'{e}r's theorem} 

For completeness, let us briefly recall here the standard results for the case of superexponential decay of $p(x \to \infty)$.
In this case, at $N \gg 1$, large deviations of $A$ follow  a scaling known as the large deviations principle (LDP) (of course, an analogous claim holds for the $x \to -\infty$ tail and $A < \left\langle A\right\rangle$), 
\be
\label{LDP}
P_{N}\left(A\right) \sim e^{-NI(A/N)} \, ,
\ee
as follows from Cram\'{e}r's theorem \cite{Cramer38, Hugo2009, MS2017}. The function
\be
I(a)=-\lim_{N\to\infty}N^{-1}\ln P_{N}\left(A=aN\right)
\ee
is usually called the rate function, and may be calculated straightforwardly: It is given by the Legendre-Fenchel transform
\be
\label{LegendreFenchel}
I\left(a\right)=\max_{k\in\mathbb{R}}\left\{ ka-\lambda\left(k\right)\right\} 
\ee
 of the cumulant generating function (CGF) $\lambda(k)$ of the distribution $p(x)$:
\be
\label{SCGF}
\lambda\left(k\right)=\ln\left\langle e^{kx_{1}}\right\rangle =\ln\int e^{kx}p\left(x\right)dx
\ee
(here and below the integration is over the entire domain of $x$'s, which is $x>0$ for $p_1(x)$ and $x \in \mathbb{R}$ for $p_2(x)$).
$I(a)$  is convex, its minimum of $I(a)$ is given by $I(a_*) = 0$ where $a_* =  \left\langle x_1\right\rangle = \mu$, and its behavior of $I(a)$ around its minimum is parabolic,
\be
\label{Iparabolic}
I\left(a\right)\simeq\frac{\left(a-\mu\right)^{^{2}}}{2\sigma^{2}} \, ,
\ee
 providing a smooth matching with the Gaussian regime  \eqref{Gaussian}.
These properties of $I(a)$ are standard and well known, see e.g. \cite{Hugo2009}.

\begin{figure*}
\includegraphics[width=0.47\textwidth,clip=]{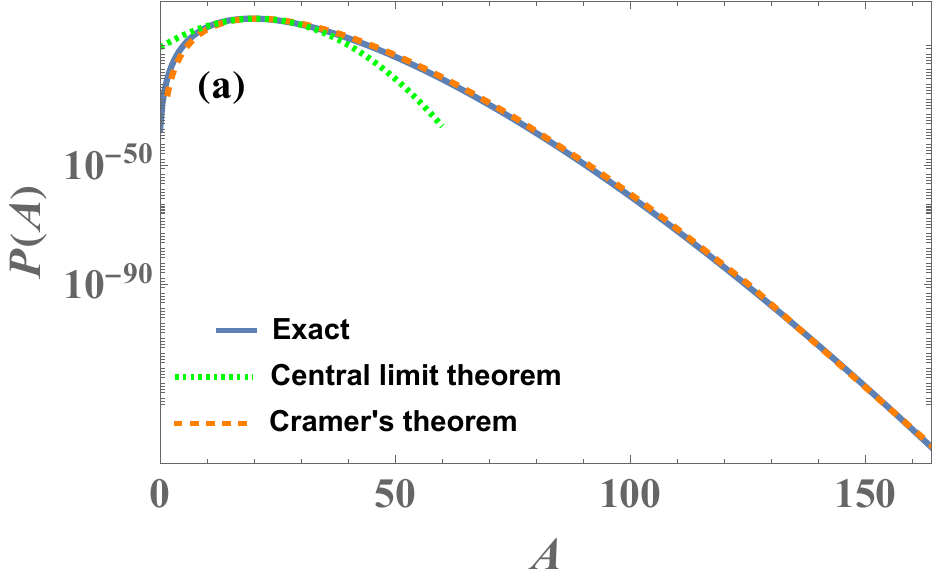}
 \hspace{2mm}
\includegraphics[width=0.47\textwidth,clip=]{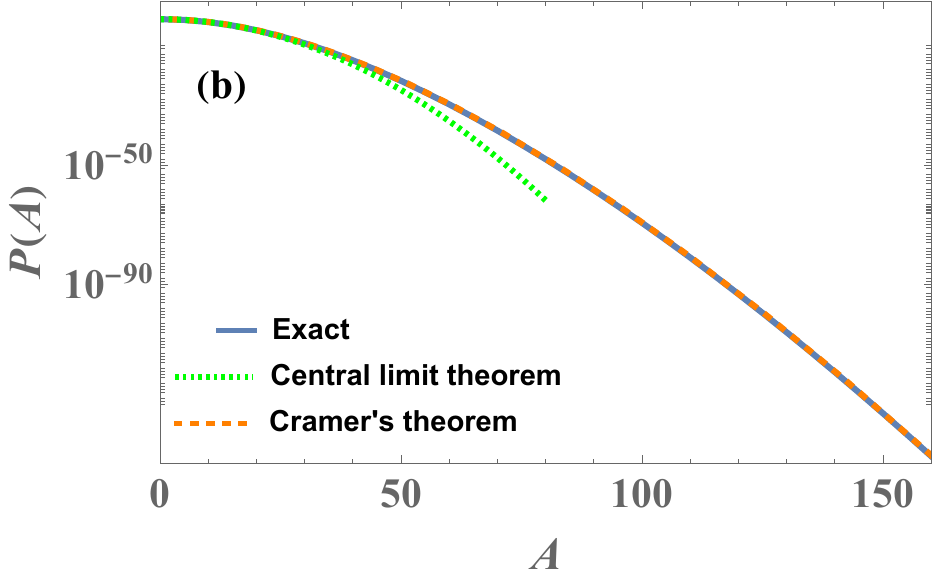}
\caption{Solid lines: Exact distribution $P_{N}\left(A\right)$ of $A = x_1 + \dots + x_N$, obtained through a numerical computation, plotted on a semi-logarithmic scale.
Dotted lines: Gaussian distribution \eqref{Gaussian} predicted by the central limit theorem.
Dashed lines: Prediction of the large deviation principle \eqref{LDP} from Cram\'{e}r's theorem, where the rate function $I(a)$ is calculated as described in the text. In the LDP prediction, we include a multiplicative factor $1/\sqrt{2\pi N \sigma^2}$ to ensure its (approximate) normalization.
Results are plotted for $N=31$ for the asymmetric \eqref{SE1} and symmetric \eqref{SE2} streched-exponential distributions of the $x_i$'s with $\alpha = 3/2$ (corresponding to superexponential decay of the tails of $p(x)$) in (a) and (b) respectively.
In (b), due to the mirror symmetry of the distribution, $P_{N}\left(A\right) = P_N(-A)$, only the right half ($A>0$) of the distribution is plotted.}
\label{figAlphaThreeHalves}
\end{figure*}

For the stretched-exponential distributions $p_1(x)$ and $p_2(x)$ with general values of $\alpha$, one can in general obtain $\lambda(k)$ by a numerical computation of the integral \eqref{SCGF} (in some particular cases, for rational $\alpha$'s, this integral may be analytically solvable). Since $\lambda(k)$ is differentiable and convex, the Legendre-Fenchel transform \eqref{LegendreFenchel} reduces to the Legendre transform \cite{TouchetteNutshell},
%\bea
%\label{Legendre}
%a&=&\lambda'\left(k\right)=\frac{\int xe^{kx}p\left(x\right)dx}{\ln\int e^{kx}p\left(x\right)dx}\,,\\
%I&=&ka-\lambda\,.
%\eea
\be
\label{Legendre}
\begin{cases}
a=\lambda'\left(k\right)=\frac{\int xe^{kx}p\left(x\right)dx}{\ln\int e^{kx}p\left(x\right)dx}\,,\\[1mm]
I=ka-\lambda\,.
\end{cases}
\ee
Eq.~\eqref{Legendre} is a parametric representation of the rate function, $a=a(k)$, $I=I(k)$, which is also straightforward to compute numerically.

Rather similarly to the CLT, also for Cram\'{e}r's theorem one usually observes a very good agreement between the prediction \eqref{LDP} and the exact $P_{N}\left(A\right)$ already at moderately large values of $N$. This may be seen in Fig.~\ref{figAlphaThreeHalves} for $p_1(x)$ and $p_2(x)$ with $\alpha = 3/2$ for $N=31$.  Note that, in the figure, we included a  multiplicative factor of $1/\sqrt{2\pi N \sigma^2}$ to the theoretical prediction \eqref{LDP}. This ensures the (approximate) normalization of the predicted PDF, and also provides a smooth matching with the Gaussian regime \eqref{Gaussian} including the pre-exponential factor, in the typical-fluctuations regime $A-\left\langle A\right\rangle \sim \sqrt{N}$.
In the large-deviations regime $A-\left\langle A\right\rangle \sim N$, there are expected to be other contributions to the pre-exponential factor, which we do not attempt to calculate here.

\subsection{Large deviations for subexponential decay $0<\alpha<1$:  Big-jump principle and condensation} 

For the one-sided case \eqref{SE1}, $p_1(x) = 0$ for $x<0$ and therefore the $x \to -\infty$ tail trivially decays superexponentially, and thus Cram\'{e}r's theorem is valid for the left tail $A< \langle A \rangle$ of $P_{N}\left(A\right)$.
However, for $0 < \alpha < 1$, $p(x \to \infty)$ decays slower than exponentially (in both cases \eqref{SE1} and \eqref{SE2}), so Cram\'{e}r's theorem breaks down as does the large deviations principle \eqref{LDP} for the right tail $A> \langle A \rangle$.
Instead, the tail $A \to \infty$ of $P_{N}\left(A\right)$ is described by the ``big-jump principle" (BJP) \cite{Chistyakov64,Foss13,Denisov08,Geluk09,Clusel06, BCV10,BUV14,VBB19, WVBB19,Gradenigo13,Barkai20, MKB98, BBBJ2000, EH05, MEZ05, EMZ06, Majumdar10, CC12, ZCG, Corberi15, BVB24}
\be
\label{BJP}
P_{N}\left(A\right) \simeq N p\left(A\right) \, ,\qquad A \to \infty.
\ee
In the context of random walks, this describes a situation in which a large deviation of the position $A$ of the random walker after $N$ steps originates in a single, unusually large step (or big jump) $x_i \simeq A$, whereas the other $N-1$ steps are all of typical size, which is negligible compared to $A$.
The factor $N$ in \eqref{BJP} is due to the $N$ different possibilities for the index $i$ at which the big jump can occur.

In between the Gaussian regime \eqref{Gaussian} and BJP regime \eqref{BJP}, there is a highly nontrivial intermediate regime, which is the main focus of the current work.
In the intermediate regime,
%If $p\left(x\to\infty\right)$ decays as a power law, the behavior in the intermediate regime is a smooth crossover.
%However, if $p(x \to \infty)$ decays as a stretched exponential,
%\be
%p\left(x\right)\sim e^{-cx^{\alpha}}
%\ee
%with $0 < \alpha < 1$ and $c>0$, then there is a highly nontrivial intermediate regime in which $\Delta A = A - \left\langle A\right\rangle $ is described by an LDP with anomalous scaling,
$P_{N}\left(A\right)$ is described by an LDP with anomalous scaling,
\be
\label{anomalousScaling}
P_{N}(A)\sim e^{-N^{\beta}f\left(\Delta A/N^{\gamma}\right)}
\ee
where $\Delta A = A - \left\langle A\right\rangle = A-N\mu$, and with scaling exponents
\be
\label{betagamma}
\beta = \alpha/\left(2-\alpha\right), \quad \gamma = 1/\left(2-\alpha\right)
\ee
that are anomalous in the sense that they are not both equal to one (as one has in the standard LDP). Moreover, the large-deviations function (LDF) is given by
\bea
\label{fdef}
f\left(y\right)&=&\min_{0 \le z \le y}F\left(y,z\right) \, ,\\
\label{Fdef}
F\left(y,z\right)&=&z^{\alpha}+\frac{\left(y-z\right)^{2}}{2\sigma^{2}} \, .
\eea
 This form of $F(y,z)$ describes $N-1$ small jumps, whose sum follows a Gaussian distribution, and a single ``big jump'', see a detailed derivation in Sec.~\ref{sec:leadingDerivation} below. The minimization in \eqref{fdef} corresponds to an optimization over the size of the big jump.
Quite remarkably, $f(y)$ has a corner singularity at a critical point $y=y_c$ at which its first derivative jumps. This singularity is often interpreted as a ``condensation transition'', which separates between a subcritical homogeneous phase $0 < y < y_c$ in which $x_1, \dots, x_N$ are all of the same order of magnitude,
and a supercritical condensed phase $y>y_c$ in which one of the $x_i$'s is of order $N^\gamma$, while the rest of the $x_i$'s are of the same order magnitude as each other.

In the subcritical regime, the LDF is exactly parabolic, $f(y) = y^2 / (2\sigma^2)$, since the minimizer in \eqref{fdef} is $z=0$. This provides a smooth matching between the intermediate regime \eqref{anomalousScaling} and the Gaussian regime \eqref{Gaussian}.
Furthermore, at $y \gg 1$ the LDF behaves as
$f(y) \simeq y ^\alpha$, since the minimizer in \eqref{fdef} is $z \simeq y$, and this provides a smooth matching between the intermediate regime \eqref{anomalousScaling} and the BJP regime \eqref{BJP}.
These results for the intermediate regime, and the properties of $f(y)$ described here have been obtained in several previous works in more general setttings \cite{EMZ06, Majumdar10, GM19, BKLP20, GIL21, GIL21b,MGM21}, but to make this paper self contained, we derive them in Section \ref{sec:leadingDerivation}, where we also give additional details.

\begin{figure*}
\includegraphics[width=0.47\textwidth,clip=]{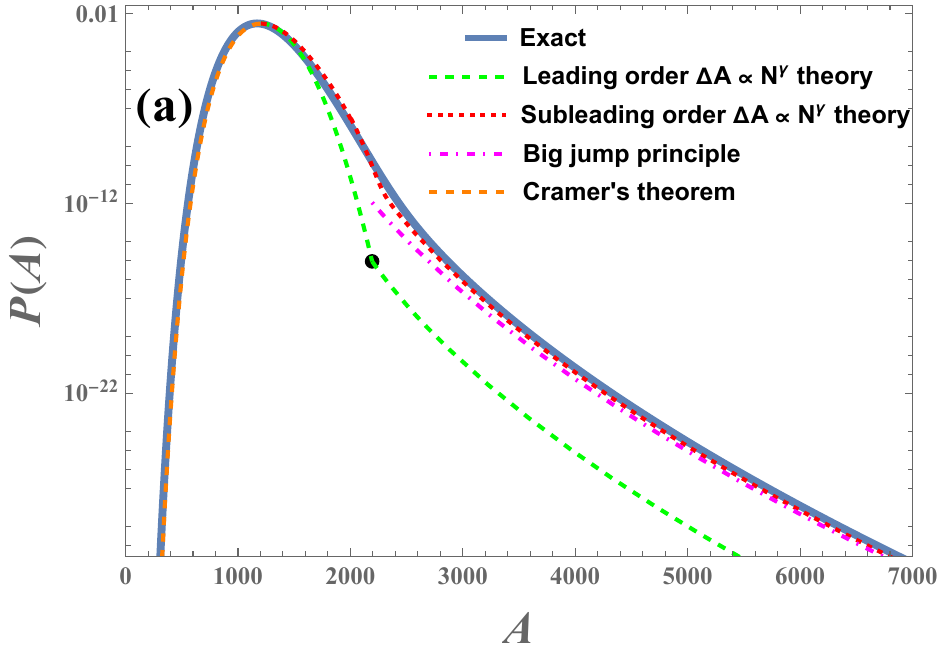}
 \hspace{2mm}
\includegraphics[width=0.47\textwidth,clip=]{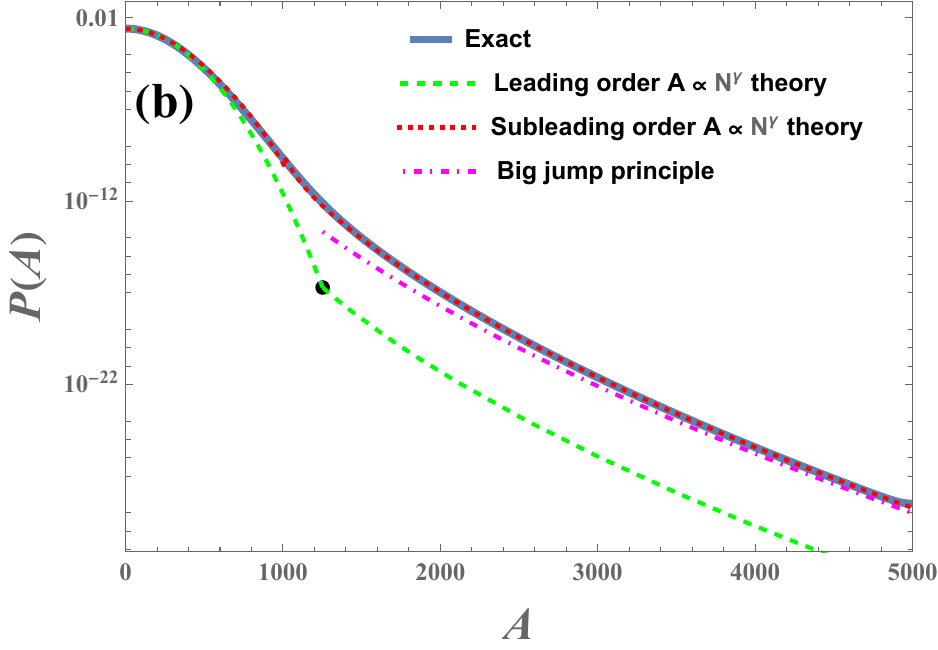}
\caption{
Solid lines: Exact distribution $P_{N}\left(A\right)$ of $A = x_1 + \dots + x_N$, obtained through a numerical computation, plotted on a semi-logarithmic scale.
Dashed lines: Leading-order prediction \eqref{anomalousScaling} for the intermediate regime $\Delta A =y N^\gamma$. The fat dot marks the critical point $y = y_c$, at which the large-deviation function $f(y)$ has a corner singularity.
Dotted lines: Subleading-order predictions, given by Eqs.~\eqref{ISeriesy}, \eqref{PNsolCondensed} and \eqref{PNsolTransition} in the homogeneous, condensed, and transition regimes respectively.
Dot-dashed lines: The prediction of the big-jump principle \eqref{BJP}.
In (a) the prediction \eqref{LDP} of Cram\'{e}r's theorem is plotted at $A < \left\langle A\right\rangle $ with an additional dashed line.
%
%Dotted lines: Gaussian distribution \eqref{Gaussian} predicted by the central limit theorem.
%Dashed lines: Prediction of the large deviation principle \eqref{LDP} from Cram\'{e}r's theorem, where the rate function $I(a)$ is calculated as described in the text. In the LDP prediction, we include a multiplicative factor $1/\sqrt{2\pi N \sigma^2}$ to ensure its (approximate) normalization.
%
Results are plotted  for $N=201$ for the asymmetric \eqref{SE1} and symmetric \eqref{SE2} streched-exponential distributions of the $x_i$'s with $\alpha = 1/2$ (corresponding to subexponential decay of the tails of $p(x)$) in (a) and (b) respectively.
In (b), due to the mirror symmetry of the distribution, $P_{N}\left(A\right) = P_N(-A)$, only the right half ($A>0$) of the distribution is plotted.
%
%
%Distribution $P_N(A)$ of the sum of $N=201$ i.i.d. random variables, each sampled from a (a) one-sided \eqref{SE1} or (b) two-sided \eqref{SE2} stretched-exponential distribution with $\alpha = 1/2$.
%Solid lines are exact numerical computations,
%Dashed lines correspond to the leading-order prediction for the intermediate regime \eqref{anomalousScaling}
In the leading-order prediction for the intermediate regime \eqref{anomalousScaling} and the LDP regime \eqref{LDP} in (a), we include a multiplicative factor $1/\sqrt{2\pi N \sigma^2}$ to ensure their (approximate) normalization.}
\label{figAlphaHalf}
\end{figure*}

Although Eq.~\eqref{anomalousScaling} becomes exact in the limit $N \to \infty$, one finds that at moderately large values of $N$, it performs rather poorly, especially in the vicinity of the critical point. This is demonstrated for the stretched-exponential distributions \eqref{SE1} and \eqref{SE2} with $\alpha=1/2$ and $N=201$ in Fig.~\ref{figAlphaHalf}. Indeed, one observes in both cases that the prediction \eqref{anomalousScaling} and the exact $P_N(A)$ differ by several orders of magnitude around the critical point and in the supercritical regime (however, in the latter regime, the BJP provides good accuracy).
This discrepancy may be of great practical importance, since it can occur for events whose probabilities could be sufficiently large to occur in realistic observations or experiments (say, of order $10^{-10}$ or higher), and the theoretical prediction \eqref{anomalousScaling} severely underestimates their likelihood.
Furthermore, this phenomenon is not unique to the specific examples of $p(x)$ considered here; It is known to happen in other cases where this type of condensation transition occurs \cite{GM19, MGM21} (and as we argue below, we expect this phenomenon to occur quite generally).

%The above formulas give a comprehensive description of typical fluctuations and large deviations of sums of $N \gg 1$ i.i.d. random variables. However, empirically one finds that, while the CLT, LDP and BJP are valid at moderately large values of $N$ ($N \simeq 30$ is usually sufficient), the anomalous scaling regime only gives a good description of $P_{N}\left(A\right)$ at extremely large $N$, of order $N \sim 10^5$ or even larger \cite{???}.

\section{Leading-order $N \gg 1$ theory in the intermediate regime for $0 < \alpha < 1$}
\label{sec:leadingDerivation}

For completeness, we begin by deriving the leading-order result \eqref{anomalousScaling}. A standard way to derive this result is to calculate the Fourier transform of $P_{N}\left(A\right)$ and then to apply the saddle-point approximation when inverting the transform, see e.g.~\cite{MGM21}. Here we present an alternative derivation in which the physics behind the condensation phenomenon is transparent from the calculation itself.
Next, we recall some properties of the large-deviation function $f(y)$ and of the related function $F(y,z)$.
Importantly, the derivations of this section will prove useful for the purpose of later calculating the subleading corrections (which are the main result of this paper) in Section \ref{sec:subleading}.

\subsection{Derivation of the anomalous-scaling large-deviations principle \eqref{anomalousScaling}}

%
%For simplicity, let us assume that $\left\langle x_{i}\right\rangle =0$. If this assumption does not hold, we define $\Delta x_i = x_i - \left\langle x_{i}\right\rangle$ and perform the analysis on $\Delta A=A-\left\langle A\right\rangle =\sum_{i=1}^{N}\Delta x_{i}$, for which we have $\left\langle \Delta x_{i}\right\rangle =0$.

We begin by writing $A$ in the form $A = x_1 + \sum_{i=2}^N x_N$, from which it is clear that $P_N(A)$ is given by the convolution
\be
\label{convolutionx1A}
P_{N}\left(A\right)=\int p\left(x_{1}\right)P_{N-1}\left(A-x_{1}\right)dx_{1} \, .
\ee
In the intermediate regime $0 < \Delta A \sim N^\gamma$ (where $\gamma$ is given in Eq.~\eqref{betagamma}), as described in the previous section, the dominant contribution to $P_N(A)$ may come from realizations of $x_1, \dots, x_N$ in which one of the $x_i$'s is very large, $x_i \sim A$, while all of the others are of typical size, and thus their sum approximately follows the Gaussian distribution \eqref{Gaussian} (we make these assumptions now and justify them aposteriori).
Using these assumptions in \eqref{convolutionx1A}, and assuming furthermore that the large summand $x_i$ is the first one, $i=1$ (and then compensating by adding a factor of $N$ since in fact any one of the summands may be the large one), we obtain
\be
\label{convolutionx1Aapprox}
P_{N}\left(A\right)\simeq N\int p\left(x_{1}\right)\frac{1}{\sqrt{2\pi N\sigma^{2}}}e^{-\left(\Delta A-x_{1}\right)^{2}/\left(2N\sigma^{2}\right)}dx_{1} \, .
\ee
Note that when using the formula \eqref{Gaussian} to approximate $P_{N-1}(A-x_1)$, we replaced $N-1$ by $N$ (the error due to this approximation is negligible at $N \gg 1$).
Plugging the stretched-exponential form \eqref{SE1} or \eqref{SE2} of $p(x_1)$ into \eqref{convolutionx1Aapprox}, and ignoring the pre-exponential factors, we obtain
\be
\label{convolutionx1ALeading}
P_{N}\left(A\right) \sim \int_{0}^{A}e^{-x_{1}^{\alpha}}e^{-\left(\Delta A-x_{1}\right)^{2}/\left(2N\sigma^{2}\right)}dx_{1} \, ,
\ee
where the integration limits are introduced since the contribution of the integration $x_1 \notin [0,A]$ is subleading.
Rescaling
\be
\label{yzdef}
\Delta A=yN^{\gamma},\quad x_{1}=zN^{\gamma} ,
\ee
and ignoring again pre-exponential terms, Eq.~\eqref{convolutionx1ALeading} becomes
\be
\label{convolutionx1ALeading2}
P_{N}\left(A=\mu N+yN^{\gamma}\right)\sim\int_{0}^{y}e^{-N^{\alpha\gamma}z^{\alpha}}e^{-N^{2\gamma-1}\left(y-z\right)^{2}/\left(2\sigma^{2}\right)}dz=\int_{0}^{y}e^{-N^{\beta}F\left(y,z\right)}dz\,,
\ee
where we used the definitions of $\beta$, $\gamma$ and $F(y,z)$ from Eqs.~\eqref{betagamma} and \eqref{Fdef}, respectively.
Finally, using the (leading-order) saddle-point approximation to evaluate the integral \eqref{convolutionx1ALeading2} (while exploiting the large parameter $N^\beta \gg 1$), the integration over $z$ is replaced by a minimization of $F(y,z)$ over $z$, which yields the leading-order result Eq.~\eqref{anomalousScaling}.

Note that, in this derivation, the physical meaning of the value of $z_{\min}$ that is the minimizer in \eqref{fdef} becomes clear.
In the subcritical phase $0 < y < y_c$, the minimizer in \eqref{fdef} is $z_{\min}=0$, describing a homogeneous phase in which all summands are of the same order of magnitude. In the supercritical phase $y > y_c$, the minimizer is nonzero $z_{\min}>0$, describing a ``condensation'' onto a single $x_i$, i.e.,
 $P_N(A)$ is dominated by realizations of $x_1, \dots, x_N$ in which one of the summands large and approximately equal to $x_i \simeq z_{\min} N^\gamma$, with the other $N-1$ summands contributing the remainder.

 One might then ask if this remainder, which is given by $A - A_i \simeq (y - z_{\min}) N^\gamma$, might not itself include a contribution from another big jump. However, if one repeats the above analysis in order to take into account this possibility, one finds that such a scenario is always sub-optimal, since the remainder is always in the subcritical regime, i.e., $0 < y - z_{\min} < y_c$ (as we show at the end of the following subsection). This implies that
%One can check that $0 < y - z_{\min} < y_c$, implying that 
the contribution of the $N-1$ remaining summands is small enough so that they are in the homogeneous phase. This justifies, aposteriori, the Gaussian approximation \eqref{Gaussian} that we used for $P_{N-1}(A-x_1)$, and physically, it means that scenarios with two (or more) big jumps give a contribution that is exponentially small and can therefore be neglected.

\subsection{Properties of the large-deviation function $f(y)$ and of $F(y,z)$}

\begin{figure*}
\includegraphics[width=0.47\textwidth,clip=]{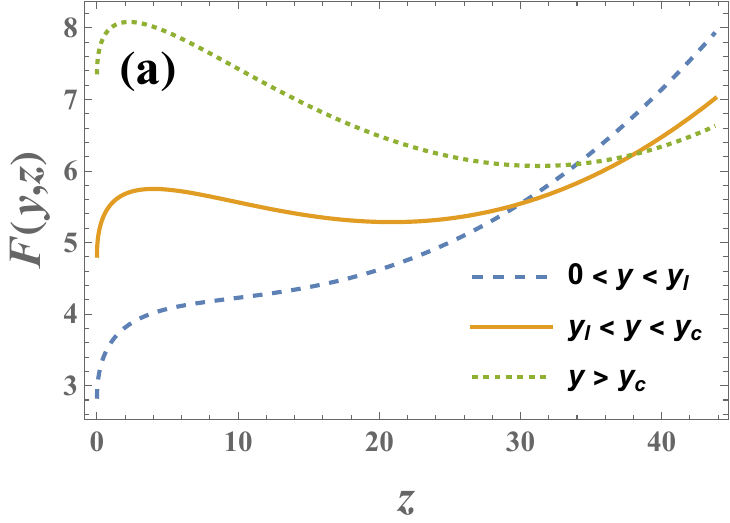}
 \hspace{2mm}
\includegraphics[width=0.47\textwidth,clip=]{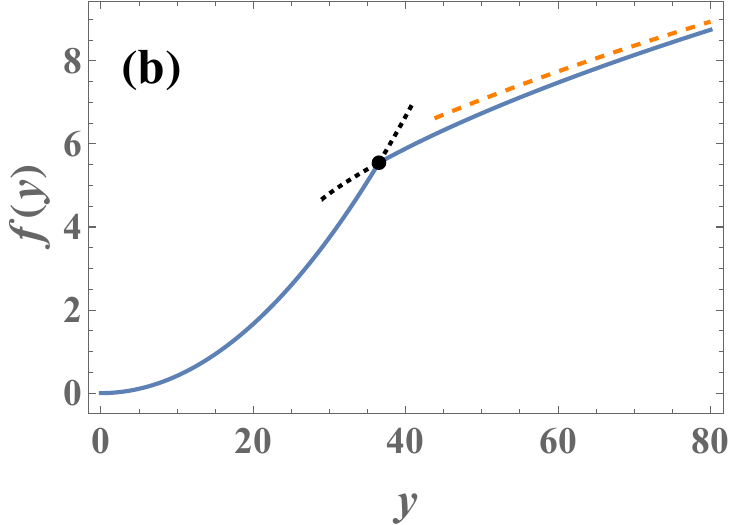}
\caption{$F(y,z)$ as a function of $z$ (a), and the large-deviation function $f(y)$ (b) that describe the intermediate regime through Eq.~\eqref{anomalousScaling}, for the symmetric case \eqref{SE2} with $\alpha=1/2$.
In (a), $F(y,z)$ is plotted for three values of $y$: One in the range $0 < y < y_l$, for which the only local minimum is at $z=0$ (dashed line), a second in the range $y_l < y < y_c$, for which an additional local minimum exists at $z_* \ne 0$ but the global minimum is at $z=0$ (solid line), and a third in the suprecritical regime $y>y_c$ for which the minimum at $z=z_*$ is the global minimum (dotted line).
The three values used in (a) are $y=26,34,42$, respectively, while $y_l = 28.9647\dots$ and $y_c = 36.4932\dots$.
In (b), the solid line corresponds to $f(y)$, the dotted lines are the continuations of the two branches of $f(y)$ to the ranges in which they are not optimal, and the dashed line is the leading-order asymptotic behavior $f(y \gg 1) \simeq y^\alpha$.
The fat dot corresponds to the critical point $y=y_c$, and the dotted line for the second branch $f_2(y)$ begins at the point $y=y_l < y_c$.}
\label{figFandf}
\end{figure*}

Here, for completeness, we recall some properties of $f(y)$ and of $F(y,z)$, defined in Eqs.~\eqref{fdef} and \eqref{Fdef}, respectively.
Clearly from Eq.~\eqref{fdef}, it is of interest to analyze the minima of $F(y,z)$ as a function of $z\in [0,y]$ for fixed $y > 0$. In Fig.~\ref{figFandf}(a), $F(y,z)$ is plotted for different fixed values of $y$ for the symmetric case \eqref{SE2} with $\alpha=1/2$.
For any $y>0$, $z=0$ is a local minimum of $F(y,z)$, at which it takes the value 
\be
f_1(y) = F(y,0) = y^2 / (2 \sigma^2) \, ,
\ee
which corresponds to one branch of the large-deviation function $f(y)$.
At $y > y_l$, where $y_l$ will be found shortly, $F(y,z)$ has another local minimum at a nonzero value $z_*$ which is the largest nonzero solution to the equation
\be
\partial_{z}F\left(y,z_{*}\right)=\alpha z_{*}^{\alpha-1}-\frac{y-z_{*}}{\sigma^{2}}=0 \, .
\ee
This equation is difficult to solve explicitly for $z_*$ in general, but it is useful to solve it for $y$:
\be
\label{yofz}
y=z_{*}+\sigma^{2}\alpha z_{*}^{\alpha-1} \, .
\ee
At $z=z_*$, $F$ takes the value
\be
\label{f2ofz}
f_{2}=F\left(y,z_{*}\right)=z_{*}^{\alpha}+\frac{\alpha^{2}\sigma^{2}z_{*}^{2\left(\alpha-1\right)}}{2}
\ee
The last two equations yield the second branch $f_2(y)$ of the large-deviation function in a parametric form.

The second branch exists only for $y > y_l$, where $y_l$ is determined from the condition
\be
\left.\frac{\partial^{2}F\left(y_{l},z\right)}{\partial z^{2}}\right|_{z=z_{l}\equiv z_{*}\left(y_{l}\right)}=\alpha\left(\alpha-1\right)z_{*}\left(y_{l}\right)^{\alpha-2}+\frac{1}{\sigma^{2}}=0\,.
\ee
leading to
\be
\label{zlSol}
z_{l}=z_{*}\left(y_{l}\right)=\left[\alpha\left(1-\alpha\right)\sigma^{2}\right]^{1/\left(2-\alpha\right)} \, ,
\ee
which, after plugging into \eqref{yofz}, yields
\be
y_l = \frac{3-\alpha}{1-\alpha}\left[\alpha\left(1-\alpha\right)\sigma^{2}\right]^{1/\left(2-\alpha\right)} \, .
\ee
This local minimum becomes the \emph{global} minimum at $y>y_{c}$, where the critical point $y_c$ is found by the condition
\be
\label{f1f2}
f_1(y_c) = f_2(y_c).
\ee
 It is useful to write Eq.~\eqref{f1f2} in terms of the corresponding value of $z_c = z_*(y_c)$, for which the equation becomes
\be
\frac{\left(z_{c}+2\sigma^{2}\alpha z_{c}^{\alpha-1}\right)^{2}}{2\sigma^{2}}=z_{c}^{\alpha}+2\sigma^{2}\alpha^{2}z_{c}^{2\left(\alpha-1\right)}\, .
\ee
Solving for $z_c$, we find
\be
z_{c}=\left[2\sigma^{2}\left(1-\alpha\right)\right]^{1/\left(2-\alpha\right)}\, ,
\ee
and plugging this into \eqref{yofz} we obtain
\be
y_c = \frac{2-\alpha}{2\left(1-\alpha\right)}\left[2\sigma^{2}\left(1-\alpha\right)\right]^{1/\left(2-\alpha\right)} \, .
\ee

The rate function $f(y)$ is given by the minimum of the two branches,
\be
f\left(y\right)=\begin{cases}
f_1(y) = y^{2}/\left(2\sigma^{2}\right)\,, & 0<y<y_{c}\,,\\[1mm]
f_{2}\left(y\right)\,, & y>y_{c}\,.
\end{cases}
\ee
$f(y)$ has a corner singularity (i.e., a discontinuity of its derivative $df/dy$) at the critical point $y=y_c$, and this singularity is normally interpreted as a first-order dynamical phase transition \cite{EMZ06, Majumdar10, GM19, BKLP20, GIL21, GIL21b,MGM21}.
Finally, at $y \gg 1$, one obtains from Eqs.~\eqref{yofz} and \eqref{f2ofz} that
\be
f(y\gg1) \simeq y^{\alpha}-\frac{\alpha^{2}\sigma^{2}y^{2\left(\alpha-1\right)}}{2} \, .
\ee
The leading order of this asymptotic behavior, $f(y\gg1) \simeq y^{\alpha}$, matches smoothly with the leading order of the BJP \eqref{BJP}.
In Fig.~\ref{figFandf}(b), $f(y)$ is plotted for the symmetric case \eqref{SE2} with $\alpha=1/2$, together with its asymptotic behavior at $y \gg 1$.

 We conclude this subsection by showing that $0 < y - z_* < y_c$, as claimed at the end of the previous subsection.
We first note that from Eq.~\eqref{yofz} it follows that $y_{2}\left(z_{*}\right)\equiv y-z_{*}$ is a monotonically decreasing function of $z_*$. It is therefore sufficient to prove that $0 < y - z_* < y_c$ for the minimal value of $z_*$, i.e., for $z_* = z_l$. Indeed, by plugging Eq.~\eqref{zlSol} into \eqref{yofz}, we find that
$y_{2}\left(z_{*}=z_{l}\right)/y_{c}=2\left(\alpha/2\right)^{1/\left(2-\alpha\right)} \! /\left(2-\alpha\right)$,
which is indeed between $0$ and $1$ for all $0 < \alpha < 1$ (as shown in  supplementary Fig.~\ref{figAlphaCoeff}).

\section{Subleading-order $N \gg 1$ theory in the intermediate regime for $0 < \alpha < 1$}
\label{sec:subleading}

%\subsection{Derivation of the subleading-order corrections}

As mentioned above, Eq.~\eqref{anomalousScaling} performs rather poorly as an approximation for $P_N(A)$ at moderately large values of $N$ (see Fig.~\ref{figAlphaHalf}) and we therefore now calculate subleading corrections. In the calculation presented in the previous subsection, one can identify three origins for such corrections:
(i) Corrections to the Gaussian distribution \eqref{Gaussian} which we used to approximate $P_{N-1}(A-x_1)$ when moving from Eq.~\eqref{convolutionx1A} to \eqref{convolutionx1Aapprox}.
(ii) Pre-exponential terms that were neglected throughout the calculation up to \eqref{convolutionx1ALeading2}.
(iii) In the condensed phase, one must take into account the subleading order corrections to the saddle-point approximation that we applied when calculating the integral in \eqref{convolutionx1ALeading2} to obtain \eqref{anomalousScaling}.
(iv) Near the transition, the contributions to $P_N(A)$ from the homogeneous and condensed scenarios are both of the same order of magnitude, and both of them must be taken into account.
We now derive the subleading corrections, beginning from the homogeneous phase, which is a little simpler since less of these correction terms are present.

\subsection{Homogeneous phase $y < y_c$}

In the homogeneous phase, corrections to the Gaussian behavior \eqref{Gaussian} are not directly related to the subexponential decay of $p(x)$. Instead, one can attribute the corrections to the lowest nonvanishing cumulants of $p(x)$, besides the mean and variance. In the typical fluctuations regime $\Delta A \sim \sqrt{N}$, these corrections are relatively small, and are given by the Edgeworth expansion \cite{Edgeworth1905, Edgeworth1906, Kendall1948}.

At larger values of $\Delta A$, these corrections become more important. 
To understand these corrections, it is useful to temporarily analyze their behavior in the absence of condensation, i.e., for $p(x \to \infty)$ that decays superexponentially (i.e., for $\alpha > 1$). In that case, large deviations of $A$ are described by the LDP \eqref{LDP}, and for intermediate values $\sqrt{N} \ll \Delta A \ll N$, there are corrections to the Gaussian distribution \eqref{Gaussian} due to higher-order corrections to the parabolic behavior \eqref{Iparabolic} of $I(a)$ around its minimum:
\be
\label{ISeries}
I\left(a\right)=\frac{\left(a-\mu\right)^{^{2}}}{2\sigma^{2}}+I_{3}\left(a-\mu\right)^{3}+I_{4}\left(a-\mu\right)^{4}+\dots.
\ee
Importantly, $I_n$ depends only on the $n$ lowest cumulants of $p(x)$. Indeed, writing the CGF as a power series
\be
\label{lambdaSeries}
\lambda\left(k\right)=\sum_{n=0}^{\infty}\lambda_{n}k^{n} \, ,
\ee
where $n! \lambda_n$ is the $n$ cumulant of $p(x)$, one then finds that the Legendre transform \eqref{Legendre} can be worked out order by order to obtain the coefficients $I_n$.
Indeed, taking the derivative of \eqref{lambdaSeries}, we obtain
\be
\label{aSeries}
a=\lambda_{1}+2\lambda_{2}k+3\lambda_{3}k^{2}+4\lambda_{4}k^{3} + \dots
\ee
Inverting Eq.~\eqref{aSeries}, we obtain
\be
\label{kSeries}
k=\frac{a-\lambda_{1}}{2\lambda_{2}}-\frac{3\lambda_{3}}{8\lambda_{2}^{3}}\left(a-\lambda_{1}\right)^{2}+\left(\frac{9\lambda_{3}^{2}}{16\lambda_{2}^{5}}-\frac{\lambda_{4}}{4\lambda_{2}^{4}}\right)\left(a-\lambda_{1}\right)^{3}+ \dots
\ee
which in turn is integrated to obtain Eq.~\eqref{ISeries} with
\be
\label{I3I4oflmabdas}
I_{3}=-\frac{\lambda_{3}}{8\lambda_{2}^{3}} \, ,\qquad I_{4}=\frac{9\lambda_{3}^{2}}{64\lambda_{2}^{5}}-\frac{\lambda_{4}}{16\lambda_{2}^{4}} \, ,
\ee
where we used that $\lambda_1 = \mu$, $\lambda_2 = \sigma^2 /2$ and $I(a=\mu)=0$. In Appendix \ref{app:highOrderCramer} analogous formulas are given up to sixth order, including expressions for $I_5$ and $I_6$ in terms of $\lambda_1, \dots, \lambda_6$.
%\be
%\label{ISeries2}
%I\left(a\right)=\frac{\left(a-\lambda_{1}\right)^{2}}{4\lambda_{2}}-\frac{\lambda_{3}}{8\lambda_{2}^{3}}\left(a-\lambda_{1}\right)^{3}+\left(\frac{9\lambda_{3}^{2}}{64\lambda_{2}^{5}}-\frac{\lambda_{4}}{16\lambda_{2}^{4}}\right)\left(a-\lambda_{1}\right)^{4}+O\left(\left(a-\lambda_{1}\right)^{5}\right) \, .
%\ee
Finally, the lowest coefficients $\lambda_n$ are obtained by calculating the lowest cumulants $\left\langle x^{n}\right\rangle_c = \lambda_n / n!$ of $p(x)$  (the first and second cumulants, $\mu$ and $\sigma^2$, respectively, were calculated above). One way to do this is to calculate the moments,
\be
\left\langle x^{n}\right\rangle =\int x^{n}p_{i}\left(x\right)dx=\begin{cases}
\frac{\Gamma\left(\frac{1+n}{\alpha}\right)}{\Gamma\left(\frac{1}{\alpha}\right)}\,, & i=2\text{ or }\left(i=1\text{ and }n\text{ is even}\right) ,\\[2mm]
0\,, & i=1\text{ and }n\text{ is odd} \, ,
\end{cases}
\ee
and then to express the cumulants through the moments, see e.g. \cite{wikiCumulants}.

Let us now return to the case of subexponential decay of $p(x \to \infty)$, and in particular to the stretched-exponential cases \eqref{SE1} or \eqref{SE2} with $0 < \alpha < 1$. In this case, $\lambda(k)$ does not exist for $k>0$ (which would correspond to fluctuations with $a > \mu$), since $\left\langle e^{kx_{1}}\right\rangle $ diverges for all $k>0$.
However, all of the cumulants of $p(x)$ are finite (one way to see this is that all of the moments $\left\langle x_{1}^n \right\rangle $, $n=0,1,2,\dots$ are finite, and one can express the cumulants through the moments). Thus, $\lambda(k)$ may be written as the formal power series \eqref{lambdaSeries} with coefficients that can be calculated explicitly, and whose radius of convergence must be zero since $\lambda(k>0)$ does not exist. In turn, one can write $I(a)$ in the form of the series \eqref{ISeries} with coefficients $I_n$ which are all finite and can be calculated explicitly. The radius of convergence of  the series \eqref{ISeries} is also expected to be zero.

Importantly, when plugging the power-series expansion \eqref{ISeries} of the rate function $I(a)$ into the LDP \eqref{LDP}, in the intermediate regime $\Delta A = y N^\gamma$ the result takes the form
\be
\label{ISeriesy}
P_{N}\left(A=\mu N + yN^{\gamma}\right)\simeq P_{N}^{\text{(hom)}}\left(A\right) = \frac{1}{\sqrt{2\pi N\sigma^{2}}}\exp\left\{ -N\left[\frac{\left(yN^{\gamma-1}\right)^{2}}{2\sigma^{2}}+I_{3}\left(yN^{\gamma-1}\right)^{3}+I_{4}\left(yN^{\gamma-1}\right)^{4}+\dots\right]\right\} \,.
\ee
In Eq.~\eqref{ISeriesy} we have included the (approximate) normalization factor from the Gaussian part of the distribution, which is recovered by keeping only the term $\propto y^2$ in the series in the exponential.
%The series in the exponential in \eqref{ISeriesy}
This series is expected to diverge for any positive value of $y$, similarly to the divergence of \eqref{ISeries} at $a > \mu$. However, one can observe that (at $N \gg 1$), since $\gamma < 1$, the first terms in the series in \eqref{ISeriesy} initially get smaller and smaller (before the series eventually blows up). This suggests that Eq.~\eqref{ISeriesy} correctly describes the homogeneous phase, as long as it the series is interpreted as an \emph{asymptotic} series. In other words, although Cram\'{e}r's theorem breaks down, its perturbative version (where the perturbative parameter is $k$ or equivalently $a - \mu$) gives a result that is correct in the homogeneous regime, to arbitrary (finite) order in perturbation theory.

This indeed appears to be the case, as we now argue theoretically and verify numerically below.
For any $L \in \mathbb{R}$, one can truncate the distribution $p(x)$ so that it vanishes at $x \ge L$, i.e., define
\be
p^{\left(L\right)}\left(x\right)=\mathcal{N}_{L}p\left(x\right)\theta\left(L-x\right) \, ,
\ee
where $\mathcal{N}_L$ is a normalization constant.
At $L \gg 1$, $p^{\left(L\right)}\left(x\right)$ differs from $p(x)$ only in the tail $x \to \infty$ (and the normalization constant $\mathcal{N}_L \simeq 1$).
In the homogeneous regime, fluctuations of $A$ are dominated by realizations of $x_1, \dots, x_N$ which are all relatively close to their typical values, and therefore the tail behavior of $p(x \to \infty)$ should not directly affect $P_N(A)$ in this regime. Therefore $P_N(A)$ must behave similarly for both $p(x)$ and $p^{\left(L\right)}\left(x\right)$ (in the homogeneous phase).
However, the truncated distribution $p^{\left(L\right)}\left(x\right)$ trivially decays superexponentially at $L \to \infty$, and thus Cram\'{e}r's theorem is applicable to it. The cumulants of $p^{\left(L\right)}\left(x\right)$ depend on $L$ of course. However, in the large $L$ limit, they converge to those of $p(x)$, and therefore, so does the behavior of the associated rate function around its minimum.

We find excellent agreement between the prediction \eqref{ISeriesy} (where the series is truncated at some finite cutoff) and the numerically-computed exact $P_N(A)$ for the stretched-exponential distributions $p(x)$,  at reasonably large values of $N$. This is demonstrated in Fig.~\ref{figAlphaHalf} for $\alpha = 1/2$ and $N=201$, where we kept the terms in the series in \eqref{ISeriesy} up to (and including) the term proportional to $y^6$  see footnote%
\footnote{see also the supplementary Fig.~\ref{figAlphaHalfTransitionRegime} for a zoom in on the transition regime for $N=201$ and an analogous figure for $N=2001$.}.
In supplementary Fig.~\ref{figAlphaHalfN51}, analogous results are plotted for $N=51$, and the agreement there is not as good, thus demonstrating the limitations of the theory.

Importantly, there are different possible qualitative behaviors of the terms in the series in \eqref{ISeriesy} that give the leading corrections to the Gaussian distribution \eqref{Gaussian} depending  on $p(x)$, as we now describe.
For $p(x)$ given by the one-sided exponential distribution \eqref{SE1}, the leading correction to the Gaussian distribution is given by the term cubic in $y$, i.e.,
\be
\label{I3correction}
\frac{P_{N}\left(A=\mu N + yN^{\gamma}\right)}{P_{N}^{\left(\text{Gauss}\right)}\left(A\right)}\simeq e^{-I_{3}y^{3}N^{3\gamma-2}}\,.
\ee
At $N \gg 1$, if $\gamma < 2/3$ this correction term is negligible, i.e., the right-hand side of \eqref{I3correction} is very close to unity, corresponding to $\alpha < \alpha_c = 1/2$, but the correction behaves as a stretched exponential if $\alpha > 1/2$ (i.e., $\gamma > 2/3$).
On the otherhand, for $p(x)$ given by the two-sided exponential distribution \eqref{SE2}, due to the mirror symmetry $p(x) = p(-x)$, the odd cumulants of $p(x)$ all vanish, so $\lambda_n$ vanishes for all odd $n$. As a result also $I_n$ vanishes for all odd $n$, so in Eq.~\eqref{ISeriesy} the leading correction to the Gaussian distribution \eqref{Gaussian} is given by the term quartic in $y$, i.e.,
\be
\label{I4correction}
\frac{P_{N}\left(A=\mu N + yN^{\gamma}\right)}{P_{N}^{\left(\text{Gauss}\right)}\left(A\right)}\simeq e^{-I_{4}y^{4}N^{4\gamma-3}}\,.
\ee
In this case, at $N \gg 1$ the correction term is negligible if $\alpha < \alpha_c$ but behaves as a stretched exponential if $\alpha > \alpha_c$, but in this case the threshold value is $\alpha_c = 2/3$. %($\gamma < 3/4$ and $\gamma > 3/4$, respectively).
%To summarize, we find that there exists a threshold value $\alpha_c$, such that for $0 < \alpha < \alpha_c$ the Gaussian approximation \eqref{Gaussian} is accurate in the homogeneous regime $0 < y < y_c$ up to corrections that become negligible at $N \to \infty$, while for $\alpha_c < \alpha < 1$, the corrections grow with $N$, and that $\alpha_c

\begin{figure*}
\includegraphics[width=0.47\textwidth,clip=]{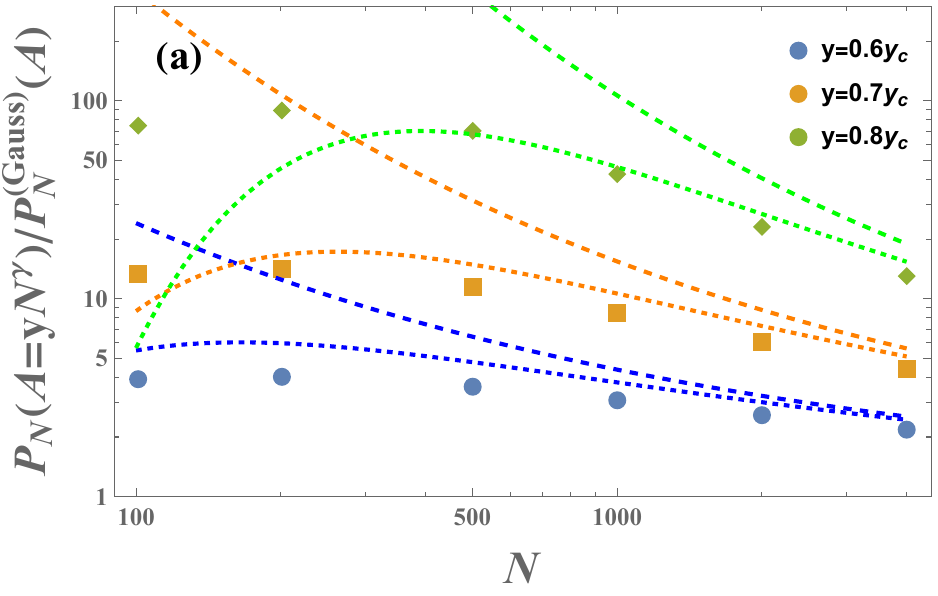}
 \hspace{2mm}
\includegraphics[width=0.47\textwidth,clip=]{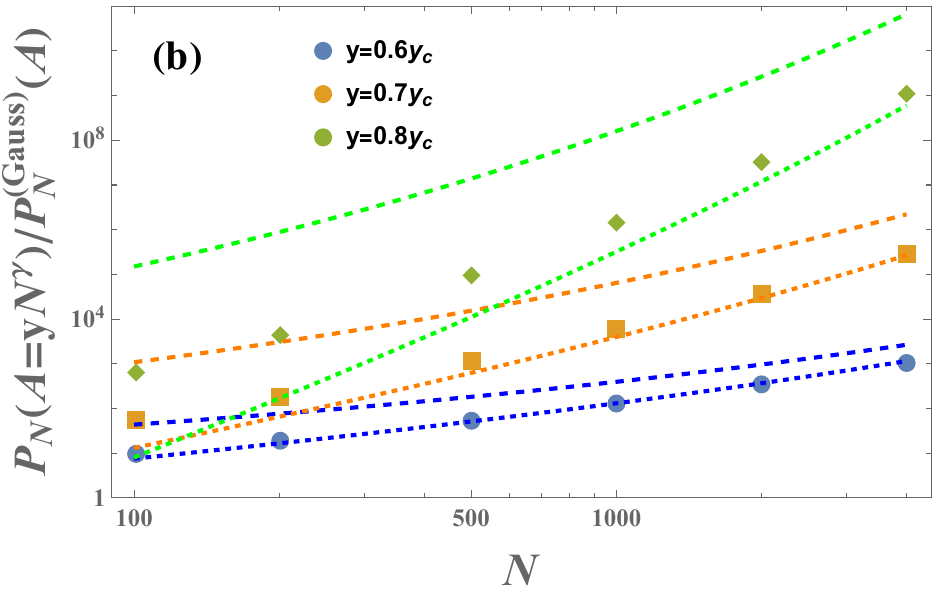}
\caption{Ratio between the prediction of our perturbative application of Cram\'{e}r's 
theorem \eqref{ISeriesy} (with the series truncated) and the Gaussian approximation \eqref{Gaussian} in the homogeneous regime $0 < y < y_c$, as a function of $N$, for $p(x)$ given by the two-sided stretched exponential distribution \eqref{SE2} with $\alpha=1/2 < \alpha_c$ (a) and $\alpha=3/4 > \alpha_c$ (b), where $\alpha_c = 2/3$ is the threshold value.
Markers represent exact results obtained from numerical computations of $P_N(A)$. Dashed lines represent the dominant correction term to the Gaussian approximation, given by the right-hand side of \eqref{I4correction}, and dotted lines represent the results obtained when including also the next nonzero term ($\propto y^6$) in the series \eqref{ISeriesy}. As seen in the figure, the correction terms decay (grow) as $N \to \infty$ for $\alpha < \alpha_c$ ($\alpha > \alpha_c$). }
\label{figMultipleNs}
\end{figure*}

In Fig.~\ref{figMultipleNs} we test Eq.~\eqref{I4correction} numerically, for $\alpha = 1/2 < \alpha_c$ and $\alpha=3/4 > \alpha_c$. In the former case, Eq.~\eqref{I4correction} becomes a small correction at $N \to \infty$, while in the latter case the correction \eqref{I4correction} grows with $N$ while higher-order corrections become negligible (see Appendix \ref{app:highOrderCramer}).
Nevertheless, as can be seen in the figure, for moderately large values $N = 100-4000$, these corrections are quite large, even for $y$ that is not very close to the critical point ($y=0.6y_c - 0.8y_c$).
This property follows from the fact that the first few values of $I_n$, for $n > 2$ are rather large, which is in turn related to a similar property of the corresponding $\lambda_n$'s. 
To give a concrete example, the kurtosis of the distribution \eqref{SE2} with $\alpha=1/2$ is
$\text{Kurt}\left(x_{i}\right)=\left\langle x_{i}^{4}\right\rangle /\left\langle x_{i}^{2}\right\rangle ^{2}=126/5$, which is fairly large (compared, e.g., to the kurtosis of the Gaussian distribution which is $3$).
The largeness of the $I_n$'s and $\lambda_n$'s is perhaps not so surprising, since the radius of convergence of the power-series
$\sum_{n=1}^{\infty} \lambda_n k^n$
and
$\sum_{n=2}^{\infty} I_n (a-\mu)^n$
are zero.
We argue that it is therefore reasonable to expect similar properties to hold in general for distributions $p(x)$ with stretched-exponential tails, leading to similar behavior as we observe for the particular examples considered here.
 We do not attempt here to find the optimal point to truncate the series \eqref{ISeriesy} in terms of maximizing the accuracy. We expect that would require a rather large number terms in the series, leading to cumbersome formulas.

As a final comment, we note that the correction terms to the Gaussian distribution, given in \eqref{ISeriesy}, match smoothly with the correction terms that one obtains from the Edgeworth expansion in the typical-fluctuations regime. This strongly supports our tacit assumption that there are no additional intermediate regimes between $\Delta A \sim \sqrt{N}$ and $\Delta A \sim N^\gamma$.
Indeed, for sufficiently small $\Delta A$, the terms in the series in \eqref{ISeriesy} are much smaller than unity, except for the first term in the series which describes the Gaussian distribution. Eq.~\eqref{ISeriesy} therefore becomes
\be
\label{ISeriesDeltaAsmall}
%P_{N}\left(A=\mu N+\Delta A\right)\simeq P_{N}^{\left(\text{Gauss}\right)}\left(A\right)\left(1-\frac{I_{3}\Delta A^{3}}{N^{2}}-\frac{I_{4}\Delta A^{4}}{N^{3}}+\dots\right)\,
P_{N}\left(A=\mu N+\Delta A\right)\simeq P_{N}^{\left(\text{Gauss}\right)}\left(A\right)\times\begin{cases}
\left(1-\frac{I_{3}\Delta A^{3}}{N^{2}}\right)\,, & I_{3}\ne0,\\[2mm]
\left(1-\frac{I_{4}\Delta A^{4}}{N^{3}}\right)\,, & I_{3}=0,
\end{cases}
\ee
describing the asymmetric and symmetric cases, respectively.
On the other hand, the Edgeworth expansion gives corrections to the typical fluctuations regime in terms of the higher cumulants \cite{EdgeworthWiki}
\be
\label{eqEdgeworth}
%P_{N}\left(A\right)\simeq P_{N}^{\left(\text{Gauss}\right)}\left(A\right)\left[1+\frac{\kappa_{3}}{3!N^{3/2}\sigma^{3}}\text{He}_{3}\left(\frac{\Delta A}{\sqrt{N}\sigma}\right)+\frac{\kappa_{4}}{4!N^{2}\sigma^{4}}\text{He}_{4}\left(\frac{\Delta A}{\sqrt{N}\sigma}\right)\right]\,,
%&& P_{N}\left(A\right)\simeq P_{N}^{\left(\text{Gauss}\right)}\left(A\right) \nn\\
%&& \times\left[1+\frac{\kappa_{3}}{3!N^{3/2}\sigma^{3}}\text{He}_{3}\left(\frac{\Delta A}{\sqrt{N}\sigma}\right)+\frac{\kappa_{4}}{4!N^{2}\sigma^{4}}\text{He}_{4}\left(\frac{\Delta A}{\sqrt{N}\sigma}\right)+\frac{\kappa_{5}}{5!N^{5/2}\sigma^{5}}\text{He}_{5}\left(\frac{\Delta A}{\sqrt{N}\sigma}\right)+\frac{\kappa_{6}+10\kappa_{3}^{2}}{6!N^{3}\sigma^{6}}\text{He}_{6}\left(\frac{\Delta A}{\sqrt{N}\sigma}\right)\right]\,,\nn\\
P_{N}\left(A\right)\simeq P_{N}^{\left(\text{Gauss}\right)}\left(A\right)\times\begin{cases}
\left[1+\frac{\kappa_{3}}{3!N^{3/2}\sigma^{3}}\text{He}_{3}\left(\frac{\Delta A}{\sqrt{N}\sigma}\right)\right]\,, & \kappa_{3}\ne0,\\[2mm]
\left[1+\frac{\kappa_{4}}{4!N^{2}\sigma^{4}}\text{He}_{4}\left(\frac{\Delta A}{\sqrt{N}\sigma}\right)\right]\,, & \kappa_{3}=0,
\end{cases}
\ee
where $\kappa_n$ is the $n$th cumulant of $A$, and $\text{He}_{n}\left(x\right)$ is the $n$th order Hermite polynomial,
$\text{He}_{3}\left(x\right)=x^{3}-3x$, $\text{He}_{4}\left(x\right)=x^{4}-6x^{2}+3$.
Using that $\kappa_n = n! N \lambda_n$, and considering the tail $\Delta A \gg \sqrt N$ of the distribution, the arguments of the Hermite polynomials become much larger than unity, so approximating $\text{He}_{n}\left(x\right) \simeq x^{n}$, Eq.~\eqref{eqEdgeworth} becomes
\be
P_{N}\left(A\right)\simeq P_{N}^{\left(\text{Gauss}\right)}\left(A\right)\times\begin{cases}
\left(1+\frac{\lambda_{3}\Delta A^{3}}{N^{2}\sigma^{6}}\right)\,, & \lambda_{3}\ne0,\\[2mm]
\left(1+\frac{\lambda_{4}\Delta A^{4}}{N^{3}\sigma^{8}}\right)\,, & \lambda_{3}=0,
\end{cases}
\ee
Which indeed coincides with \eqref{ISeriesDeltaAsmall} due to \eqref{I3I4oflmabdas}.
To summarize this point: Corrections to the central limit theorem in the homogeneous phase $y<y_c$ are given by the Edgeworth expansion in the typical fluctuations regime, and by Eq.~\eqref{ISeriesy} in the large-deviations regime, and the two formulas have a joint regime of validity $N^{\gamma} \gg \Delta A \gg \sqrt{N}$ in which their predictions coincide.

\subsection{Condensed phase $y > y_c$}

In the condensed phase, the dominant contribution to $P_N(A)$ is due to realizations in which one of the $x_i$'s is very large, of order $\Delta A$ itself, while the rest of the $x_i$'s are of similar order of magnitude.
%Let us do the derivation first for the asymmetric case \eqref{SE1}.
Our starting point is the formula \eqref{convolutionx1A} that gives $P_N(A)$ as the convolution of the distributions of $x_1$ and that of $A-x_1$. 
Following similar steps to those of the leading-order calculation, we consider the probability that $x_1$ is very large, while $x_2, \dots, x_N$ are of typical size (and compensate by adding a factor of $N$ in front of the integral, since in fact any of the $x_i$'s could have been the large one).
We thus approximate $P_{N-1}\left(A-x_{1}\right)$ by the expression \eqref{ISeriesy} from the homogeneous phase, we plug in
$p(x) = C e^{-x^\alpha}$
where $C = 1/\Gamma\left(1+\frac{1}{\alpha}\right)$
and $C = 1/2\Gamma\left(1+\frac{1}{\alpha}\right)$
for the asymmetric \eqref{SE1} and symmetric \eqref{SE2} cases, respectively,
and performing the change of variables \eqref{yzdef} we obtain
%\bea
%\label{convolutionSubleading1}
%&&P_{N}\left(A=\mu N+yN^{\gamma}\right)\simeq C\sqrt{\frac{N}{2\pi\sigma^{2}}} \nn\\
%&&\qquad\qquad \times \int_{0}^{y} \exp\left\{ -N^{\alpha/\left(2-\alpha\right)}F\left(y,z\right)-N\left[I_{3}\left(\left(y-z\right)N^{\gamma-1}\right)^{3}+I_{4}\left(\left(y-z\right)N^{\gamma-1}\right)^{4}+\dots\right]\right\} N^{\gamma}dz\,,
%\eea
\be
\label{convolutionSubleading1}
P_{N}\left(A=\mu N+yN^{\gamma}\right)\simeq C\sqrt{\frac{N}{2\pi\sigma^{2}}}\int_{0}^{y}e^{-N^{\beta}F\left(y,z\right)-N\left[I_{3}\left(\left(y-z\right)N^{\gamma-1}\right)^{3}+I_{4}\left(\left(y-z\right)N^{\gamma-1}\right)^{4}+\dots\right]}N^{\gamma}dz\,,
\ee
which is analogous to Eq.~\eqref{convolutionx1ALeading2}, but with subleading correction terms taken into account.
We are now ready to perform the saddle-point approximation to evaluate the integral in \eqref{convolutionSubleading1}.
In the leading order, the result is of course given by Eq.~\eqref{anomalousScaling}. We note that the subleading terms in the series in the exponential in \eqref{convolutionSubleading1} are much smaller than the leading-order term $\propto N^{\alpha/(2-\alpha)}$. Therefore, their effect on the value of the minimizer $z_*$ in the saddle-point approximation is subleading, and they can be safely taken out of the integral by replacing in them $z \to z_*$.
Evaluating the remaining integral to subleading order in the saddle-point approximation, we obtain
\bea
\label{PNsolCondensed}
P_{N}\left(A=\mu N+yN^{\gamma}\right)&\simeq& P_{N}^{\text{(con)}}\left(A\right) \nn\\
&=&\frac{CN^{\gamma+1/2}e^{-N\left[I_{3}\left(\left(y-z_{*}\right)N^{\gamma-1}\right)^{3}+I_{4}\left(\left(y-z_{*}\right)N^{\gamma-1}\right)^{4}+\dots\right]}}{\sqrt{2\pi\sigma^{2}}}\sqrt{\frac{2\pi}{N^{\beta}\partial_{z}^{2}F\left(y,z_{*}\right)}}e^{-N^{\beta}F\left(y,z_{*}\right)} \nn\\[1mm]
&=&\frac{Ce^{-N\left[I_{3}\left(\left(y-z_{*}\right)N^{\gamma-1}\right)^{3}+I_{4}\left(\left(y-z_{*}\right)N^{\gamma-1}\right)^{4}+\dots\right]}}{\sqrt{1-\sigma^{2}\alpha\left(1-\alpha\right)z_{*}^{\alpha-2}}}Ne^{-N^{\beta}F\left(y,z_{*}\right)} \, .
\eea
Deep into the condensed phase, at $y \gg 1$ [describing the far tail $P_N(A \to \infty)$], one has $z_* \simeq y \gg 1$,
$y - z \ll 1$ and
$F\left(y,z_{*}\right) = f(y) \simeq y^\alpha$,
so Eq.~\eqref{PNsolCondensed} simplifies to
\be
P_{N}\left(A=\mu N+yN^{\gamma}\right)\simeq CNe^{-N^{\beta}y^{\alpha}}=CNe^{-\Delta A^{\alpha}} \, ,
\ee
in agreement with the big-jump principle \eqref{BJP}.

As in the homogneous regime, here too the qualitative behavior of the correction term behaves as a power law in $N$ for $0 < \alpha < \alpha_c$ and as a stretched exponential for $\alpha_c< \alpha < 1$ (with the same values of $\alpha_c$ for the asymmetric and symmetric cases as given above).
Moreover, we find that the prediction \eqref{PNsolCondensed} shows excellent agreement with the exact $P_N(A)$ in the condensed regime, as demonstrated in Fig.~\ref{figAlphaHalf} for $\alpha = 1/2$ and $N=201$ [where we kept the terms in the series in \eqref{PNsolCondensed} up to (and including) the term proportional to $y^6$].

\subsection{Transition regime $y \simeq y_c$}

In the transition regime $y \simeq y_c$, 
one has to apply the saddle-point approximation more carefully, since the two local minima of the function $F(y,z)$ both yield contributions to $P_N(A)$ that are of similar order of magnitude. Neither of these two contributions should be neglected in the transition regime, and we therefore take both of them into account, i.e.,
%\sout{one can write $P_N(A)$ as the sum of two contributions that are of similar order of magnitude,}
\be
\label{PNsolTransition}
P_{N}\left(A\right) \simeq P_{N}^{\text{(hom)}}\left(A\right) + P_{N}^{\text{(con)}}\left(A\right).
\ee
The first term on the right-hand side of \eqref{PNsolTransition} corresponds to homogeneous realizations of $x_1, \dots, x_N$ and the second term corresponds to condensed realizations, and they are given by Eqs.~\eqref{ISeriesy} and \eqref{PNsolCondensed}, respectively (where both series must be truncated as explained above).
Note that the contribution of condensed realizations \eqref{PNsolCondensed} is defined at $y > y_l$ and it can therefore be taken into account also at subcritical values $y_l < y < y_c$. However,  Eq.~\eqref{PNsolCondensed} breaks down at $y\simeq y_l$ (and it diverges at $y=y_l$). As a result, to obtain a reliable prediction for $P_N(A)$, it is in practice best to choose an intermediate value $y_l < y_m < y_c$, and to use Eqs.~\eqref{ISeriesy} and \eqref{PNsolTransition} for $0 < y < y_m$ and $y>y_m$, respectively (at sufficiently large $N$, the difference between the two formulas becomes negligible at $y=y_m$). Indeed, this is how we produced the dotted line in Fig.~\ref{figAlphaHalf}, which is in excellent agreement with the exact $P_N(A)$ at all values of $y$.

Let us estimate the width of the transition regime. We can define the transition regime to be $A \in [A_1, A_2]$ where $A_1$ and $A_2$ are determined by the requirement that one of the two contributions (homogeneous and condensed) is twice as large as the other, i.e.,
\be
\label{A1A2def}
P_{N}^{\text{(hom)}}\left(A_1\right) = 2 P_{N}^{\text{(con)}}\left(A_1\right), \qquad P_{N}^{\text{(hom)}}\left(A_2\right) = P_{N}^{\text{(con)}}\left(A_2\right) / 2
\ee
(of course, one does not have to choose a factor of 2, any factor of order unity would work).
To calculate the width of the transition, $A_2- A_1$, it is sufficient to use the leading-order approximations for the homogeneous and condensed contributions. These are given by Eq.~\eqref{anomalousScaling} with the two branches $f_1(y)$ and $f_2(y)$ of the large-deviation function, respectively.
One finds that the width of the transition is determined by the condition
$\left(A_{2}-A_{1}\right)/N^{\gamma}\sim N^{-\beta}$ which leads to
\be
A_{2}-A_{1}\sim N^{\gamma-\beta}=N^{\left(1-\alpha\right)/\left(2-\alpha\right)} \, .
\ee
For $\alpha = 1/2$ for instance, this gives $N^{1/3}$.

\subsection{Extension to more general $p(x)$}

Our results are straightforward to extend to more general $p(x)$ with stretched-exponential tails. Indeed, let us assume, e.g., that the $x \to \infty$ tail decays as a stretched exponential with a power-law prefactor
\be
\label{pxGeneral}
p\left(x\to\infty\right)\simeq CA^{\zeta}e^{-bA^{\alpha}}
\ee
with $0 < \alpha < 1$ and $b > 0$
Then, in the intermediate regime, $P_N(A)$ is described, in the leading order, by Eq.~\eqref{anomalousScaling} where definition of the large-deviation function must be modified by replacing $F(y,z)$ from \eqref{Fdef} by
\be
F\left(y,z\right)=b z^{\alpha}+\frac{\left(y-z\right)^{2}}{2\sigma^{2}}
\ee
to account for $b$ [and of course $\mu$ and $\sigma$ must be calculated for the distribution $p(x)$].
In the subleading order, $P_N(A)$ is still described by Eq.~\eqref{ISeriesy} in the homogeneous regime [with coefficients $I_n$ that must be calculated from the cumulants of $p(x)$].

In the condensed phase, the saddle-point calculation is affected by the power-law term in \eqref{pxGeneral}. However, the effect is rather simple to take into account, since the effect of this term on the value of $z_*$ is small, and therefore Eq.~\eqref{PNsolCondensed} is simply modified by including an additional factor of $(z_*N^\gamma)^\zeta$, i.e., one obtains
\be
\label{PNsolCondensedGeneral}
P_{N}\left(A=\mu N+yN^{\gamma}\right)\simeq\frac{Ce^{-N\left[I_{3}\left(\left(y-z_{*}\right)N^{\gamma-1}\right)^{3}+I_{4}\left(\left(y-z_{*}\right)N^{\gamma-1}\right)^{4}+\dots\right]}}{\sqrt{1-\sigma^{2}\alpha\left(1-\alpha\right)}}z_{*}^{\zeta-\left(\alpha-2\right)/2}N^{1+\gamma\zeta}e^{-N^{\beta}F\left(y,z_{*}\right)}\,,
\ee
where the value of $z_* = z_*(y)$ is modified to account for $b$, i.e., Eq.~\eqref{yofz} is replaced by
\be
\label{yofzGeneral}
y=z_{*}+\sigma^{2}\alpha b z_{*}^{\alpha-1} \, .
\ee
In the transition regime, $P_N(A)$ is still given by the sum of the contributions of the formulas for homogeneous and condensed realizations, i.e., by Eq.~\eqref{PNsolTransition}.

In this analysis, we assumed not only that the $x \to \infty$ tail of $p(x)$ is given by a stretched exponential \eqref{pxGeneral}, but also that all of its cumulants are finite. The latter assumptions may not hold, e.g., if $p(x \to -\infty)$ decays as a power law \cite{Barkai20, BVB24}, and an interesting direction for future work  would be to investigate the behavior of $P_N(A)$ in such cases.

\section{Summary and discussion}
\label{sec:Discussion}

To summarize, we studied the distribution of the sum $A$ of $N$ i.i.d. random variables, each distributed according to a given distribution $p(x)$ with a stretched exponential tail at $x \to \infty$.
We calculated the subleading corrections to the (previously-known) leading-order prediction \eqref{anomalousScaling} for the large-$N$ asymptotic behavior of $P_N(A)$ in the intermediate regime $\Delta A \sim N^\gamma$.
The leading-order prediction can perform very poorly for moderately large $N$'s (in the range $10^2 - 10^4$), especially for $A$ which is not far from the critical point at which a condensation transition occurs, see e.g.~Fig.~\ref{figAlphaHalf}.
For $p(x)$ given by a pure asymmetric \eqref{SE1} or symmetric \eqref{SE2} stretched exponential, we calculated the subleading-order corrections to Eq.~\eqref{anomalousScaling}. These are given by Eqs.~\eqref{ISeriesy}, \eqref{PNsolCondensed} and \eqref{PNsolTransition} in the homogeneous, condensed, and transition regimes respectively.
It is interesting to note that despite the fact that Cram\'{e}r's theorem breaks down, the perturbative version of it which we developed here is valid and gives meaningful and correct results.
We extended our results to more general $p(x)$ with stretched-exponential tails.

We found a threshold value $\alpha_c$ such that for $\alpha < \alpha_c$, the  corrections to Eq.~\eqref{anomalousScaling} become negligible as $N \to \infty$, while for $\alpha > \alpha_c$ they grow with $N$ as a stretched exponential.
Nevertheless, at moderately large $N$, it is in practice important to take into account even some of the correction terms that would eventually decay in the limit $N \to \infty$.
We argue that the reason behind the relative largeness of the subleading corrections at moderately large $N$ is due to the rapid growth of the cumulants $\left\langle x^{n}\right\rangle _{c}$ of the distribution $p(x)$ with $n$. We expect this feature to be quite generic, i.e., to occur for any $p(x)$ which decays slower than exponentially, since the associated CGF $\lambda(k)$ diverges at any $k>0$.
 The results of the current paper are not rigorously proven at the level of mathematical theorems, and it would be useful to perform a more rigorous treatment of the problem that we studied.

Finally, we restricted ourselves here to the case of sums of i.i.d.~random variables. However, a breakdown of the large-deviation principle \eqref{LDP}, replaced instead by an intermediate regime with anomalous scaling as in \eqref{anomalousScaling} and accompanied by a condensation transition, can occur for other systems as well. Examples include sums of correlated random variables, e.g., linear statistics of eigenvalues of random matrices \cite{NMV10, VMS24}. Moreover, anomalous scaling has also been observed in distributions of long-time averages in continuous-time dynamics for several systems \cite{NT18, DH2019, MeersonGaussian19, BH20, Smith22OU, NT22, SmithMajumdar22, Smith24Absx, Ferre25, VM25}, with tails described by instantons that are analogous to the big-jump principle \cite{BVBB25}. There too, the leading-order theory does not perform very well in the intermediate regime for moderately-long averaging times \cite{Smith22OU, Smith24Absx}. Furthermore, it would be interesting to explore whether condensation transitions accompanied by anomalously-scaled LDP's can occur in other types of systems too.
 One example is large deviations of time-averaged observables in deterministic, chaotic dynamical systems, which have recently attracted renewed interest \cite{Smith22Chaos, Monthus23Chaos, RCP23, Lippolis24, Monthus25Pelikan, DS25}.
A subleading-order theory analogous to the one developed here could facilitate such studies, since it enables testing the theory for moderately-large averaging times, for which numerical results can be obtained. This is in contrast to the huge averaging times that are necessary to test the leading-order theory, for which it is more difficult to obtain numerical results (see however \cite{SmithFiniteDifferences24} for some recent progress for a particular continuous-time system).

%\cite{HT09, NMV10, NT18, MeersonGaussian19, GM19, DH2019, JackHarris20, BH20, BKLP20, MLMS21, MGM21, GIL21, GIL21b, Smith22OU, NT22, SmithMajumdar22}

\bigskip

\section*{Acknowledgments}

I thank Baruch Meerson and Satya Majumdar for useful discussions and advice.
I acknowledge support from the Israel Science Foundation (ISF) through Grant No. 2651/23, and from the Golda Meir Fellowship.

\bigskip

\begin{appendices}

\section{Higher-order perturbative Cram\'{e}r calculations}
\label{app:highOrderCramer}
\renewcommand{\theequation}{A\arabic{equation}}
\setcounter{equation}{0}

In the main text, the perturbative Cram\'{e}r calculation is carried out explicitly to yield the series expansion of the rate function $I(a)$ up to quartic order in terms of the cumulants of $p(x)$, see Eq.~\eqref{I3I4oflmabdas}. In this appendix, we give an expansion up to sixth order.

Taking the derivative of \eqref{lambdaSeries}, we obtain
\be
\label{aSeries6}
a=\lambda_{1}+2\lambda_{2}k+3\lambda_{3}k^{2}+4\lambda_{4}k^{3} +5\lambda_{5}k^{4} + 6\lambda_{6}k^{5} +  \dots
\ee
Inverting Eq.~\eqref{aSeries6}, we obtain
\bea
\label{kSeries6}
k&=&\frac{a-\lambda_{1}}{2\lambda_{2}}-\frac{3\lambda_{3}}{8\lambda_{2}^{3}}\left(a-\lambda_{1}\right)^{2}+\frac{9\lambda_{3}^{2}-4\lambda_{2}\lambda_{4}}{16\lambda_{2}^{5}}\left(a-\lambda_{1}\right)^{3}-\frac{5\left(27\lambda_{3}^{3}-24\lambda_{2}\lambda_{4}\lambda_{3}+4\lambda_{2}^{2}\lambda_{5}\right)}{128\lambda_{2}^{7}}\left(a-\lambda_{1}\right)^{4}\nn\\
&-&\frac{3\left(-189\lambda_{3}^{4}+252\lambda_{2}\lambda_{4}\lambda_{3}^{2}-60\lambda_{2}^{2}\lambda_{5}\lambda_{3}-32\lambda_{2}^{2}\lambda_{4}^{2}+8\lambda_{2}^{3}\lambda_{6}\right)}{256\lambda_{2}^{9}}\left(a-\lambda_{1}\right)^{5} + \dots
\eea
which in turn is integrated to obtain Eq.~\eqref{ISeries} with $I_3$ and $I_4$ given by Eq.~\eqref{I3I4oflmabdas} of the main text, and
\be
\label{I5I6oflmabdas}
I_5 = -\frac{27\lambda_{3}^{3}}{128\lambda_{2}^{7}}+\frac{3\lambda_{4}\lambda_{3}}{16\lambda_{2}^{6}}-\frac{\lambda_{5}}{32\lambda_{2}^{5}},\qquad I_6 = \frac{189\lambda_{3}^{4}}{512\lambda_{2}^{9}}-\frac{63\lambda_{4}\lambda_{3}^{2}}{128\lambda_{2}^{8}}+\frac{15\lambda_{5}\lambda_{3}}{128\lambda_{2}^{7}}+\frac{\lambda_{4}^{2}}{16\lambda_{2}^{7}}-\frac{\lambda_{6}}{64\lambda_{2}^{6}} \, .
\ee

We now briefly discuss the relevance of the terms that arise in the series in  \eqref{ISeriesy}. The term proportional to $I_n$ in this series is
\be
I_n y^n N^{n \gamma - n+1} \, .
\ee
In the limit $N \to \infty$ (with constant $y$), this term will approach infinity if $\gamma > (n-1) / n$, or approach zero if $\gamma < (n-1) / n$. Recalling Eq.~\eqref{betagamma}, the two cases correspond to
$\alpha>\left(n-2\right)/(n-1)$
and
$\alpha<\left(n-2\right)/(n-1)$
respectively (as explained in the main text for the particular cases $n=3,4$).
It follows that, for a given $\alpha$, the result \eqref{ISeriesy} is correct, up to corrections that are relatively small, if the series is truncated at
\be
\label{nmax}
n_{\max}=\left\lfloor \left(2-\alpha\right)/\left(1-\alpha\right)\right\rfloor
\ee
where
$\left\lfloor \cdots\right\rfloor $
denotes the integer part.
Thus, the larger $\alpha$  is (i.e., the closer it is to 1), the more terms should be taken in the series in \eqref{ISeriesy}.
For $\alpha=1/2$ and $\alpha=3/4$, for which the correction terms are plotted in Fig.~\ref{figMultipleNs}, Eq.~\eqref{nmax} yields
$n_{\max} = 3$ and $n_{\max} = 5$, respectively (one must recall of course that the figure corresponds to the symmetric case \eqref{SE1} for which $I_n$ vanishes for all odd $n$).
However, as shown in the figure, at moderately large $N$, we observe that some corrections may still be relatively large (before eventually becoming negligible at $N \to \infty$), so for practical purposes it may be useful to take into account a few additional correction terms that are of higher order than $n_{\max}$, as we did, e.g., in Fig.~\ref{figAlphaHalf}.

\bigskip

\newpage

\section{Supplementary figures}
\label{app:suppFigs}
\renewcommand{\theequation}{B\arabic{equation}}
\setcounter{equation}{0}

\begin{figure*}
\includegraphics[width=0.4\textwidth,clip=]{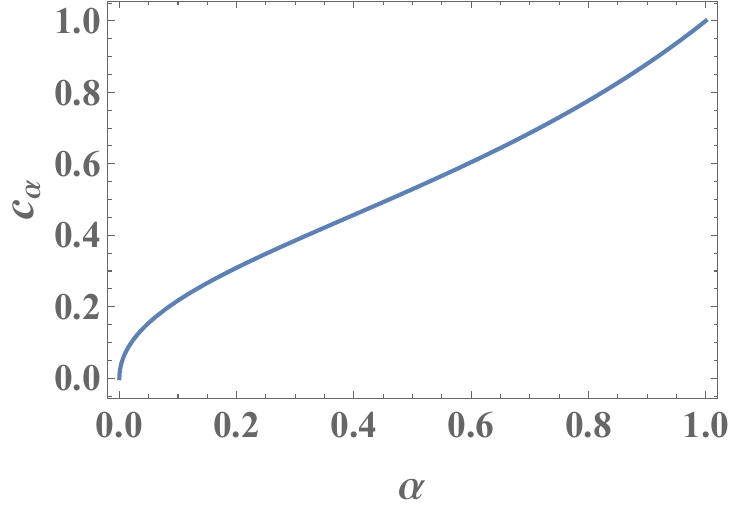}
\caption{The coefficient
$c_\alpha = y_{2}\left(z_{*}=z_{l}\right)/y_{c}=2\left(\alpha/2\right)^{1/\left(2-\alpha\right)}/\left(2-\alpha\right)$
as a function of $\alpha$.
As seen in the figure, $0 < c_\alpha < 1$ for all $0 < \alpha < 1$.}
\label{figAlphaCoeff}
\end{figure*}

\begin{figure*}
\includegraphics[width=0.47\textwidth,clip=]{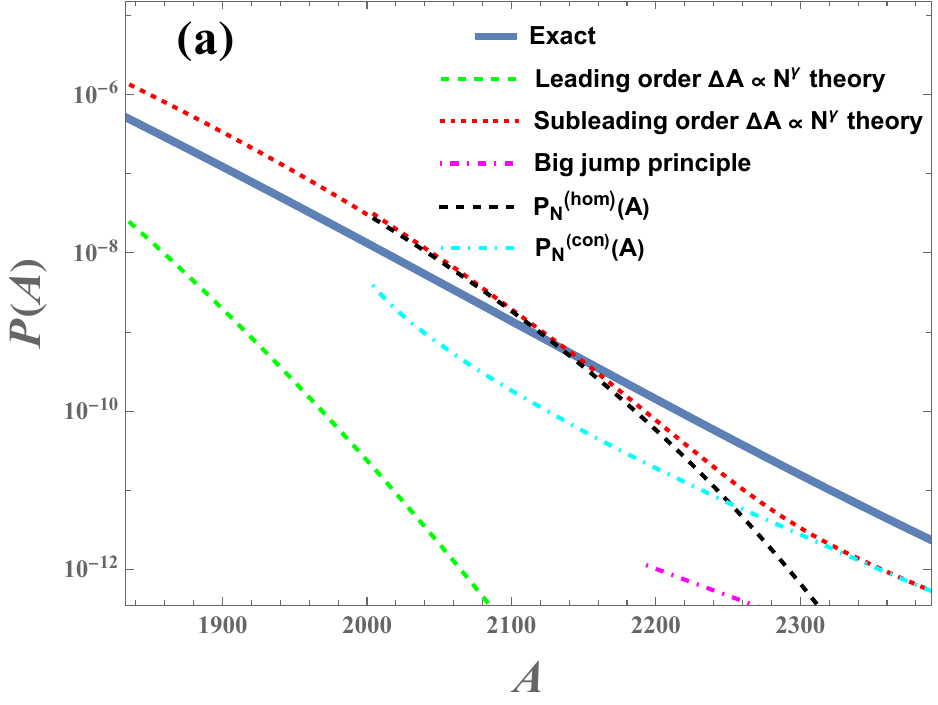}
 \hspace{2mm}
\includegraphics[width=0.47\textwidth,clip=]{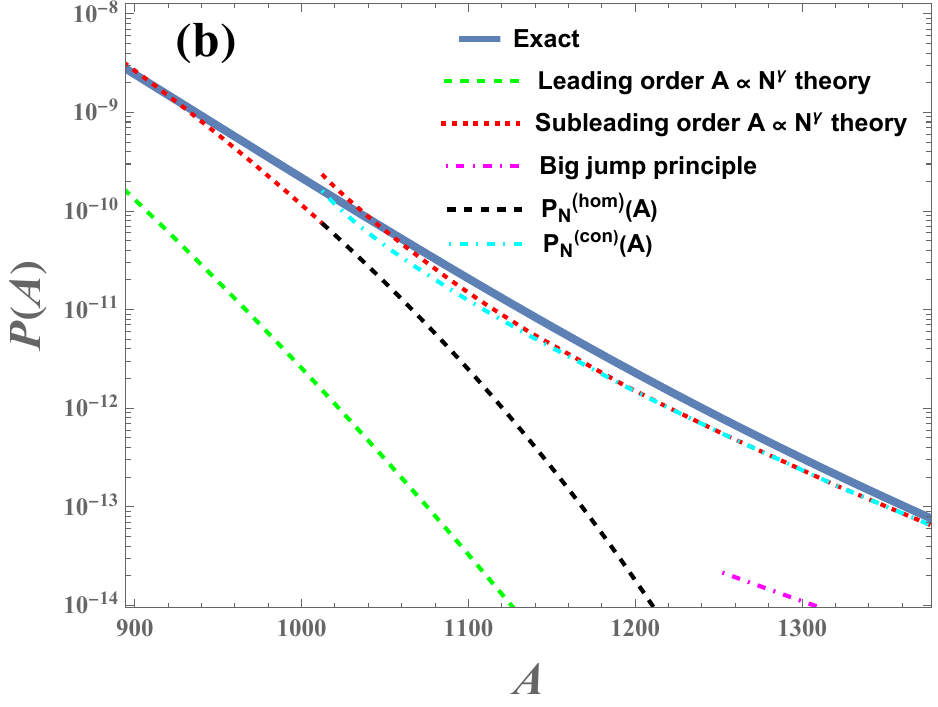}
\includegraphics[width=0.47\textwidth,clip=]{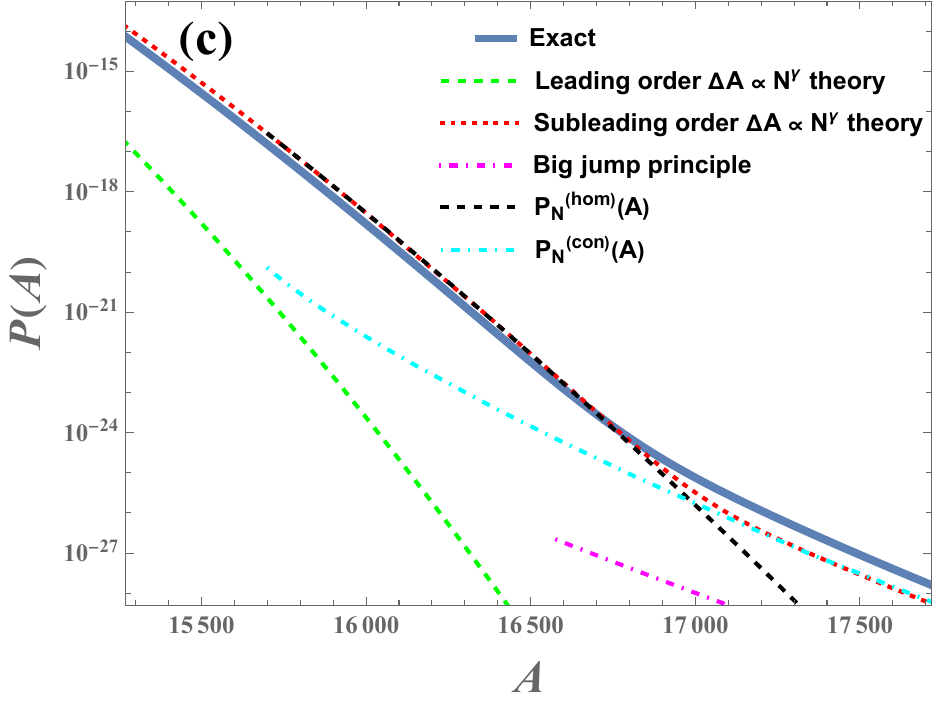}
 \hspace{2mm}
\includegraphics[width=0.47\textwidth,clip=]{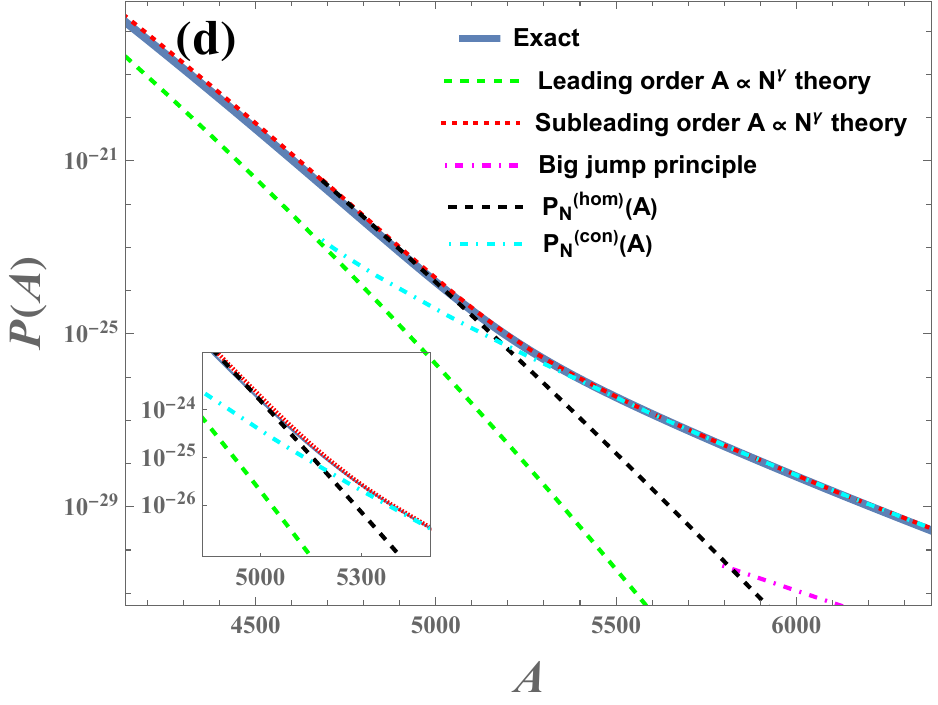}
\caption{ 
(a) and (b): A zoom in of Fig.~\ref{figAlphaHalf} on the transition regime (for $N=201$). Also plotted are the individual contributions
$P_{N}^{\text{(hom)}}\left(A\right)$ and $P_{N}^{\text{(con)}}\left(A\right)$
of the homogeneous and condensed scenarios at $y>y_m$, respectively, see Eq.~\eqref{PNsolTransition} (at $y<y_m$, the contribution of the condensed scenario is neglected, so that only the homogeneous contribution remains).
In (c) and (d), analogous plots are shown for a larger value of $N=2001$. One can observe that overall, the agreement between the theoretical predictions and the exact (numerical) result improves as $N$ is increased.
In particular, for the symmetric case, one observes in (d) good agreement between Eq.~\eqref{PNsolTransition} and the exact $P_N(A)$ in the regime where the contributions $P_{N}^{\text{(hom)}}\left(A\right)$ and $P_{N}^{\text{(con)}}\left(A\right)$ are of the same order of magnitude (see inset for an even closer zoom in on this regime). For the asymmetric case, even at $N=2001$ the theory is not accurate enough to perform a similar test.}
\label{figAlphaHalfTransitionRegime}
\end{figure*}

\begin{figure*}
\includegraphics[width=0.47\textwidth,clip=]{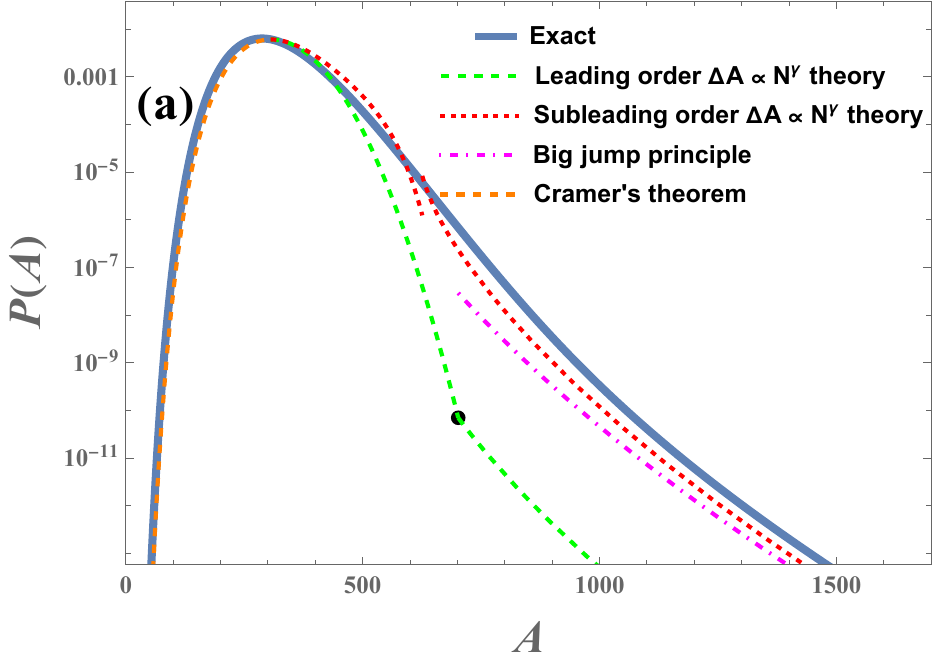}
 \hspace{2mm}
\includegraphics[width=0.47\textwidth,clip=]{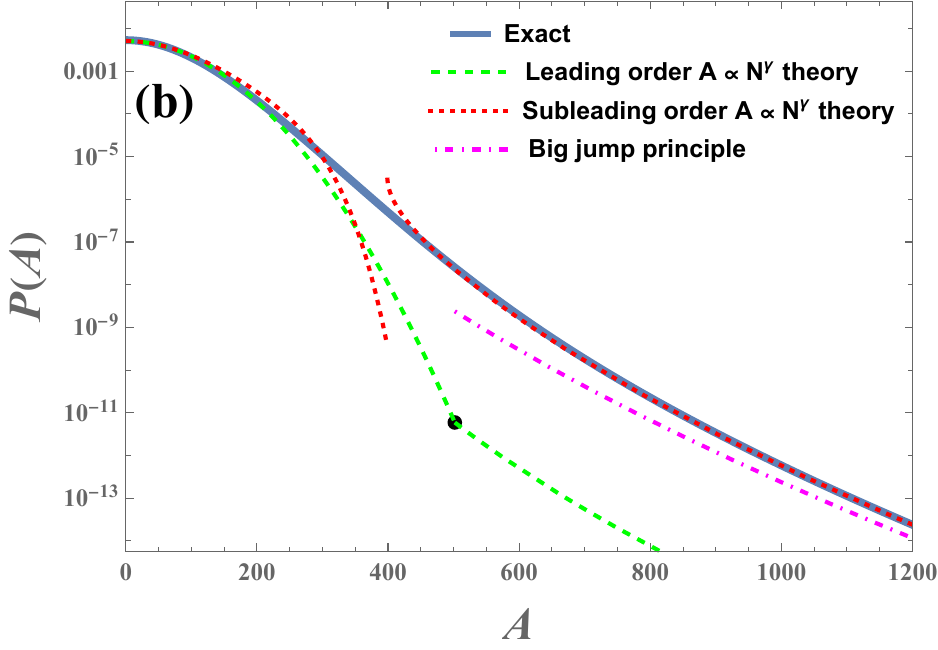}
\caption{Same as Fig.~\ref{figAlphaHalf} but for $N=51$. Here the theoretical predictions describe the exact distribution less successfully, especially in the transition regime.}
\label{figAlphaHalfN51}
\end{figure*}

 This appendix includes the supplementary Figs.~\ref{figAlphaCoeff},  \ref{figAlphaHalfTransitionRegime} and \ref{figAlphaHalfN51}, which support the findings described in the main text.

\end{appendices}

\end{document}